\documentclass[
	amsmath,
	amssymb,
	a4paper,
	aps,		
	pre,		
	reprint,	
	fleqn,
	showpacs,
	floatfix
]{revtex4-1}

\usepackage[utf8]{inputenc}
\usepackage{microtype}

\usepackage{bm}
\usepackage{amsthm}
\usepackage{braket}
\usepackage{kbordermatrix}
	\renewcommand{\kbldelim}{(}
	\renewcommand{\kbrdelim}{)}

\usepackage[lining,semibold]{libertine}
\usepackage[libertine,slantedGreek,bigdelims,vvarbb]{newtxmath}
\useosf 
\renewcommand{\Delta}{\upDelta}

\usepackage{graphicx}
\usepackage{dcolumn}
\usepackage[caption=false]{subfig}
\usepackage{color}

\usepackage{tabularx}
\usepackage{booktabs}
\usepackage{multirow}

\usepackage[version=3]{mhchem}

\usepackage{hyperref}

\newcommand{\myTitle}{Conservation Laws shape Dissipation}

\newcommand{\myName}{Riccardo Rao}

\newcommand{\myAffiliation}{Complex Systems and Statistical Mechanics, Physics and Materials Science Research Unit, University of Luxembourg, L-1511 Luxembourg, G.D.~Luxembourg}

\newcommand{\myAdvisor}{Massimiliano Esposito}

\newcommand{\de}{\mathrm{d}}
\newcommand{\dt}{\mathrm{d}_{t}}

\newcommand{\at}[2]{\left.{#1}\right|_{#2}}
\newcommand{\ave}[1]{\left\langle {#1} \right\rangle}

\newcommand{\transpose}{^{\mathrm{T}}}

\newcommand{\kb}{k_\mathrm{B}}

\DeclareMathOperator{\coker}{coker}

\DeclareMathOperator{\coim}{coim}
\DeclareMathOperator{\diag}{diag}

  {\left\lbrace\begin{array}{@{}l@{}}}%
  {\end{array}\right.}

\theoremstyle{definition}

\theoremstyle{definition}

\theoremstyle{definition}

\theoremstyle{plain}

\definecolor{butter1}{rgb}{0.98,0.91,0.31}
\definecolor{butter2}{rgb}{0.93,0.83,0}
\definecolor{butter3}{rgb}{0.77,0.63,0}
\definecolor{skyblue1}{rgb}{0.45,0.62,0.81}
\definecolor{skyblue2}{rgb}{0.2,0.39,0.64}
\definecolor{skyblue3}{rgb}{0.13,0.29,0.53}
\definecolor{scarlet1}{rgb}{0.93,0.16,0.16}
\definecolor{scarlet2}{rgb}{0.8,0,0}
\definecolor{scarlet3}{rgb}{0.64,0,0}
\definecolor{chameleon1}{rgb}{0.54,0.88,0.2}
\definecolor{chameleon2}{rgb}{0.45,0.82,0.09}
\definecolor{chameleon3}{rgb}{0.3,0.6,0.02}
\definecolor{orange1}{rgb}{0.98,0.68,0.24}
\definecolor{orange2}{rgb}{0.96,0.47,0}
\definecolor{orange3}{rgb}{0.8,0.36,0}
\definecolor{plum1}{rgb}{0.68,0.5,0.66}
\definecolor{plum2}{rgb}{0.46,0.31,0.48}
\definecolor{plum3}{rgb}{0.36,0.21,0.4}
\definecolor{chocolate1}{rgb}{0.91,0.72,0.43}
\definecolor{chocolate2}{rgb}{0.75,0.49,0.07}
\definecolor{chocolate3}{rgb}{0.56,0.35,0.01}
\definecolor{aluminium1}{rgb}{0.93,0.93,0.92}
\definecolor{aluminium2}{rgb}{0.82,0.84,0.81}
\definecolor{aluminium3}{rgb}{0.73,0.74,0.71}
\definecolor{aluminium4}{rgb}{0.53,0.54,0.52}
\definecolor{aluminium5}{rgb}{0.33,0.34,0.32}
\definecolor{aluminium6}{rgb}{0.18,0.2,0.21}  

\definecolor{webgreen}{rgb}{0,.5,0}
\definecolor{webbrown}{rgb}{.6,0,0}
\definecolor{grigio}{rgb}{.85,.85,.85} 
\definecolor{RoyalBlue}{rgb}{0.0, 0.14, 0.4}

\hypersetup{%
    colorlinks=true, linktocpage=true, pdfstartpage=1, pdfstartview=FitV,%
    breaklinks=true, pdfpagemode=UseNone, pageanchor=true, pdfpagemode=UseOutlines,%
    plainpages=false, bookmarksnumbered, bookmarksopen=true, bookmarksopenlevel=1,%
    hypertexnames=true, pdfhighlight=/O,
    urlcolor=webbrown, linkcolor=RoyalBlue, citecolor=webgreen, 
    pdftitle={\myTitle},%
    pdfauthor={\textcopyright\ \myName},%
    pdfsubject={},%
    pdfkeywords={},%
    pdfcreator={pdfLaTeX},%
    pdfproducer={LaTeX REVTeX}%
}

\newcommand{\st}[1]{\{#1\}}
\newcommand{\rf}{y_{\mathrm{p}}}
\newcommand{\push}{y_{\mathrm{f}}}
\newcommand{\sym}{\psi}
\newcommand{\sqLabel}{\kappa}
\newcommand{\sq}{Y^{\sqLabel}}
\newcommand{\excMtx}{\delta Y}

\newcommand{\trj}{\bm n_{t}}
\newcommand{\nof}[1]{\mathsf{N}_{#1}}

\newcommand{\avef}[1]{\langle {#1} \rangle}

\begin{document}

\title{\myTitle}

\author{\myName}
\affiliation{\myAffiliation}
\author{\myAdvisor}
\affiliation{\myAffiliation}

\date{\today. Published in \emph{New~J.~Phys.}, DOI:~\href{https://doi.org/10.1088/1367-2630/aaa15f}{10.1088/1367-2630/aaa15f}}

\begin{abstract}
	Starting from the most general formulation of stochastic thermodynamics---\emph{i.e.} a thermodynamically consistent nonautonomous stochastic dynamics describing systems in contact with several reservoirs---, we define a procedure to identify the conservative and the minimal set of nonconservative contributions in the entropy production.
	The former is expressed as the difference between changes caused by time-dependent drivings and a generalized potential difference.
	The latter is a sum over the minimal set of flux--force contributions controlling the dissipative flows across the system.   
	When the system is initially prepared at equilibrium (\emph{e.g.} by turning off drivings and forces), a finite-time detailed fluctuation theorem holds for the different contributions.
	Our approach relies on identifying the complete set of conserved quantities and can be viewed as the extension of the theory of generalized Gibbs ensembles to nonequilibrium situations.    
\end{abstract}


\maketitle

\section{Introduction}

Stochastic Thermodynamics provides a rigorous formulation of nonequilibrium thermodynamics for open systems described by Markovian dynamics \cite{sekimoto10,jarzynski11,seifert12,vandenbroeck15}.
Thermodynamic quantities fluctuate and the first and second law of thermodynamics can be formulated along single stochastic trajectories.
Most notably, entropy-production fluctuations exhibit a universal symmetry, called fluctuation theorem (FT).
This latter implies, among other things, that the average entropy production (EP) is non-negative.
Besides being conceptually new, this framework has been shown experimentally relevant in many different contexts \cite{ciliberto17}.
It also provides a solid ground to analyze energy conversion \cite{seifert12,verley14:universal,proesmans16}, the cost of information processing \cite{berut12,horowitz14,jun14,parrondo15,ouldridge17}, and speed--accuracy trade-offs \cite{rao15:proofreading,barato16} in small systems operating far from equilibrium.   

In stochastic thermodynamics, the dynamics is expressed in terms of Markovian rates describing transition probabilities per unit time between states.
The thermodynamics, on the other hand, assigns conserved quantities to each system state (\emph{e.g.} energy and particle number).
This means that transitions among states entail an exchange of these conserved quantities between the system and the reservoirs.
The core assumption providing the connection between dynamics and thermodynamics is local detailed balance.
It states that the log ratio of each forward and backward transition rate corresponds to the entropy changes in the reservoirs caused by the exchange of the conserved quantities (divided by the Boltzmann constant).
These changes are expressed as the product of the entropic intensive fields characterizing the reservoirs (\emph{e.g.} inverse temperature, chemical potential divided by temperature) and the corresponding changes of conserved quantities in the reservoirs, in accordance to the fundamental relation of equilibrium thermodynamics in the entropy representation.
Microscopically, the local detailed balance arises from the assumption that the reservoirs are at equilibrium \cite{esposito12}.

In this paper, we ask a few simple questions which lie at the heart of nonequilibrium thermodynamics.
We consider a system subject to time-dependent drivings---\emph{i.e.} nonautonomous---and in contact with multiple reservoirs.
What is the most fundamental representation of the EP for such a system?
In other words, how many independent nonconservative forces multiplied by their conjugated flux appear in the EP?
Which thermodynamic potential is extremized by the dynamics in absence of driving when the forces are set to zero?
How do generic time-dependent drivings affect the EP?
Surprisingly, up to now, no systematic procedure exists to answer these questions. 
We provide one in this paper based on a systematic identification of conserved quantities.
While some of them are obvious from the start (\emph{e.g.} energy and particle number) the others are system specific and depend on the way in which reservoirs are coupled to the system and on the topology of the network of transitions. 

The main outcome of our analysis is a rewriting of the EP, Eq.~\eqref{eq:epAwesome}, which identifies three types of contributions:
A driving contribution caused by the nonautonomous mechanisms, a change of a generalized Massieu potential, and a flow contribution made of a sum over a fundamental set of flux-force contributions.
For autonomous systems relaxing to equilibrium---all forces must be zero---, the first and the third contributions vanish and the dynamics maximizes the potential.
This amounts to a dynamical realization of the maximization of the Shannon entropy under the constrains of conserved quantities, which is commonly done by hand when deriving generalized Gibbs distributions. 
For (autonomous) steady-state dynamics, the first two contributions vanish and we recover the results of Ref.~\cite{polettini16}, showing that conservation laws reduce the number of forces created by the reservoirs.  
The key achievement of this paper is to demonstrate that conservation laws are essential to achieve a general and systematic treatment of stochastic thermodynamics. 

Important results ensue.
We show that system-specific conservation laws can cause the forces to depend on system quantities and not only on intensive fields.
We derive the most general formulation of finite-time detailed FTs expressed in terms of measurable quantities.
This result amounts to make use of conservation laws on the FT derived in Ref.~\cite{bulnescuetara14}.
We identify the minimal cost required for making a transformation from one system state to another one.
In doing so we generalize to multiple reservoirs the nonequilibrium Landauer's principle derived in Refs.~\cite{hasegawa10,takara10,esposito11}.
We also apply our method to four different models which reveal different implications of our theory.

This paper is organized as follows.
In \S~\ref{sec:st} we derive an abstract formulation of stochastic thermodynamics.
We then describe the procedure to identify all conserved quantities, which we use to rewrite the local detailed balance in terms of potential and (nonconservative) flow contributions.
In \S~\ref{sec:tep} we use the above decomposition to establish balance equations along stochastic trajectories, which allow us to formulate our finite-time detailed FT, \S~\ref{sec:ftDFR}.
In \S~\ref{sec:EA} we discuss the ensemble average description of our EP decompositions, as well as the nonequilibrium Landauer's principle.
Four detailed applications conclude our analysis in \S~\ref{sec:applications}.
The first is referenced systematically throughout the paper to illustrate our results.
It describes two quantum dots coupled to three reservoirs.
The second describes a quantum point contact tightly coupled to a quantum dot and shows that thermodynamic forces can depend on system features.
The third is a molecular motor exemplifying the differences between conservative and nonconservative forces in relation to the topology of the network used to model it.
The last one is a randomized grid illustrating that our formalism becomes essential when analyzing more complex systems.

\section{Edge Level Descpription}
\label{sec:st}

After formulating stochastic thermodynamics for continuous-time Markov jump processes from a graph-theoretical perspective, we describe the general procedure to identify conservative and nonconservative contributions to the local detailed balance.

\subsection{Stochastic Dynamics}

We consider an externally driven open system characterized by a discrete number of states, which we label by $n$.
Allowed transitions between pairs of states, $n \overset{\nu}{\leftarrow } m$, are described by directed edges, $e \equiv (nm, \nu)$.
The index $\nu = 1, \dots$ labels different types of transitions between the same pair of states, \emph{e.g.} transitions due to different reservoirs.
The time evolution of the probability of finding the system in the state $n$, $p_{n} \equiv p_{n}(t)$, is governed by the master equation
\begin{equation}
	\dt p_{n} = {\textstyle\sum_{e}} D^{n}_{e} \, \avef{J^{e}} \, ,
	\label{eq:ME}
\end{equation}
which is here written as a continuity equation.
Indeed, the \emph{incidence matrix} $D$,
\begin{equation}
	D^{n}_{e} :=
	\begin{cases}
		+ 1 & \text{if } \overset{e}{\longrightarrow} n \\
		- 1 & \text{if } \overset{e}{\longleftarrow} n \\
		  0 & \text{otherwise}
	\end{cases} \, ,
	\label{eq:incidence}
\end{equation}
associates each edge to the pair of states that it connects.
It thus encodes the \emph{network topology}.
On the graph identified by $\st{n}$ and $\st{e}$, it can be thought of as a (negative) divergence operator when acting on edge-space vectors---as in the master equation \eqref{eq:ME}---or as a gradient operator when acting on state-space vectors.
The ensemble averaged edge probability currents,
\begin{equation}
	\avef{J^{e}}
	= w_{e} p_{o(e)}
	\, ,
	\label{eq:aveCurrents}
\end{equation}
are expressed in terms of the transition rates, $\st{w_{e} \equiv w_{e}(t)}$, which describe the probability per unit time of observing a transition along the edge $e$.
The function
\begin{equation}
	o(e) := m \, , \quad \text{for } \overset{e}{\longleftarrow} m \, ,
	\label{eq:origin}
\end{equation}
maps each edge into the state from which it originates.
For thermodynamic consistency, each transition $e \equiv (nm, \nu)$ with finite rate $w_{e}$ has a corresponding backward transition $-e \equiv (mn, \nu)$ with a finite rate $w_{-e}$.
The stochastic dynamics is assumed to be ergodic at any time.

\paragraph*{Notation}
From now on, upper--lower indices and Einstein summation notation will be used:
repeated upper--lower indices implies the summation over all the allowed values for those indices.
The meaning of all the indices that will be used is summarized in Tab.~\ref{tab:indices}.
Time derivatives are denoted by ``$\dt$'' or ``$\partial_{t}$'' whereas the overdot ``$\, \dot{} \,$'' is reserved for rates of change of quantities that are not exact time derivatives.
We also take the Boltzmann constant $\kb$ equal to $1$.

\begin{table}
	\centering
	\begin{tabular}{lcr}
		\toprule
		index{ }	& label for{  } 	& number \\
		\midrule
		$n$			& state 					& $\nof{\mathrm{n}}$ \\
		$e$			& transition 				& $\nof{\mathrm{e}}$ \\
		$\sqLabel$ 	& system quantity 			& $\nof{\upkappa}$ \\
		$r$ 		& reservoir		 			& $\nof{\mathrm{r}}$ \\
		$y \equiv (\sqLabel,r)${ }			& conserved quantity $\sq$ from reservoir $r$		& $\nof{\mathrm{y}}$ \\
		$\alpha$	& cycle 				& $\nof{\upalpha}$ \\
		$\lambda$	& conservation law and conserved quantity 	& $\nof{\uplambda}$ \\
		$\rf$		& ``potential'' $y$ 	& $\nof{\uplambda}$ \\
		$\push$		& ``force'' $y$		& $\nof{\mathrm{y}} - \nof{\uplambda}$ \\
		$\rho$		& symmetry 			& $\nof{\uprho}$ \\
		$\eta$ 	& fundamental cycle 	& $\nof{\upalpha} - \nof{\uprho}$ \\
		& &  $= \nof{\mathrm{y}} - \nof{\uplambda}$ \\
		\bottomrule
	\end{tabular}
	\caption{
		Summary of the indices used throughout the paper and the object they label.
	}
	\label{tab:indices}
\end{table}

\subsection{Stochastic Thermodynamics}

\begin{figure}[t]
	\centering
	\includegraphics[width=.40\textwidth]{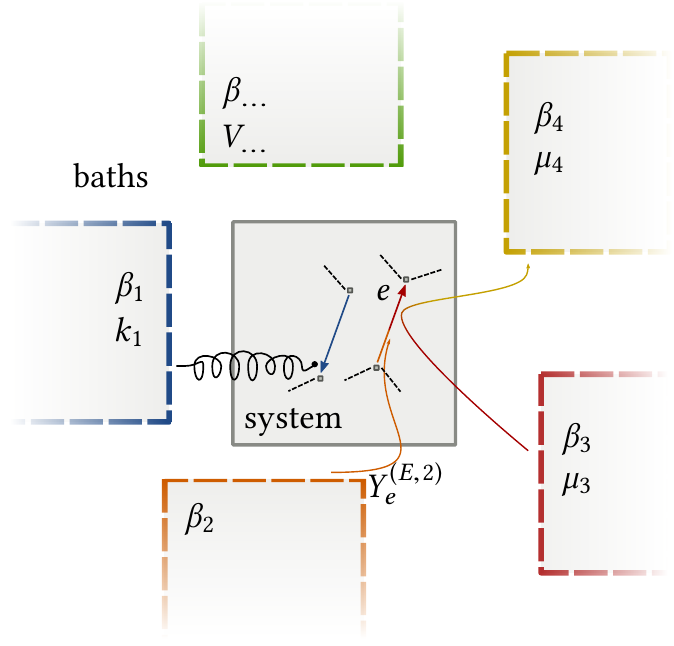}
	\caption{\label{fig:systemBath}
		Pictorial representation of a system coupled to several reservoirs.
		Transitions may involve more than one reservoir and exchange between reservoirs.
		Work reservoirs are also taken into account.
	}
\end{figure}

Physically, each system state, $n$, is characterized by given values of some \emph{system quantities}, $\st{\sq_{n}}$, for $\sqLabel=1,\dots,\nof{\upkappa}$, which encompass the internal energy, $E_{n}$, and possibly additional ones, see Tab.~\ref{tab:fYpairs} for some examples.
These must be regarded as \emph{conserved quantities} in the total system, as their change in the system is always balanced by an opposite change in the reservoirs.
Indeed, when labeling the reservoirs with $\st{r}$, for $r=1,\dots,\nof{\mathrm{r}}$, the balance equation for $\sq$ can be written as
\begin{equation}
	\underbrace{\sq_{n} - \sq_{m} \equiv \sq_{n'} D^{n'}_{e}}_{\text{system}} = {\textstyle\sum_{r}} \underbrace{\excMtx^{(\sqLabel,r)}_{e}}_{\text{reservoir }r} \, .
	\label{eq:balanceGlobal}
\end{equation}
where $\excMtx^{(\sqLabel,r)}_{e}$ quantifies the flow of $\sq$ supplied by the reservoir $r$ to the system along the transition $e$.
For the purpose of our discussion, we introduce the index $y=(\sqLabel,r)$, \emph{i.e.} \emph{the conserved quantity $\sq$ exchanged with the reservoir $r$}, and define the matrix $\excMtx$ whose entries are $\st{\excMtx^{y}_{e} \equiv \excMtx^{(\sqLabel,r)}_{e}}$.
Enforcing microscopic reversibility, one concludes that $\excMtx^{y}_{e} = - \excMtx^{y}_{-e}$.
As a first remark, more than one reservoir may be involved in each transition, see Fig.~\ref{fig:systemBath} and the application in \S~\ref{sec:QPC}.
As a second remark, the conserved quantities may not be solely $\st{\sq}$, since additional ones may arise due to the topological properties of the system, as we will see in the next subsection.

\begin{table}
	\centering
	\begin{tabular}{lr}
		\toprule
		system quantity $\sq${  } 	& intensive field $f_{(\sqLabel,r)}$ \\
		\midrule
		energy, $E_{n}$					& inverse temperature, $\beta_{r}$	\\
		particles number, $N_{n}${ }	& chemical potential, $- \beta_{r} \mu_{r}$ \\
		charge, $Q_{n}$					& electric potential, $- \beta_{r} V_{r}$	\\
		displacement, $X_{n}$			& generic force, $- \beta_{r} k_{r}$ \\
		angle, $\theta_{n}$				& torque, $- \beta_{r} \tau_{r}$ \\
		\bottomrule
	\end{tabular}
	\caption{
		Examples of system quantity--intensive field conjugated pairs in the entropy representation \cite[\S~2-3]{callen85}.
		$\beta_{r} := 1/T_{r}$ denotes the inverse temperature of the reservoir.
		Since charges are carried by particles, the conjugated pair $(Q_{n}, - \beta_{r} V_{r})$ is usually embedded in $(N_{n}, - \beta_{r} \mu_{r})$, see \emph{e.g.} Refs.~\cite{*[][{, \S~1.7.2.}] {beard08},rutten09}.
	}
	\label{tab:fYpairs}
\end{table}

Each reservoir $r$ is characterized by a set of \emph{entropic intensive fields}, $\st{f_{(\sqLabel,r)}}$ for ${\sqLabel=1,\dots,\nof{\upkappa}}$, which are conjugated to the exchange of the system quantities $\st{\sq}$ \cite[\S~2-3]{callen85}.
A short list of $\sq$--$f_{(\sqLabel,r)}$ conjugated pairs is reported in Tab.~\ref{tab:fYpairs}.
The thermodynamic consistency of the stochastic dynamics is ensured by the \emph{local detailed balance property},
\begin{equation}
	\ln \frac{w_{e}}{w_{-e}} = - f_{y} \excMtx^{y}_{e} + S_{n} D^{n}_{e} \, .
	\label{eq:ldb}
\end{equation}
It relates the log ratio of the forward and backward transition rates to the entropy change generated in the reservoirs, \emph{i.e.} minus the entropy flow $\st{- f_{y} \excMtx^{y}_{e}}$.
The second term on the rhs is the internal entropy change occurring during the transition, since $S_{n}$ denotes the internal entropy of the state $n$.
This point is further evidenced when writing the entropy balance along a transition
\begin{equation}
	\hspace{-1.4em}
	\ln \frac{w_{e} p_{o(e)}}{w_{-e} p_{o(-e)}}
	= {\textstyle\sum_{r}} \left\{ - {\textstyle\sum_{\sqLabel}} f_{(\sqLabel,r)} \excMtx^{(\sqLabel,r)}_{e} \right\} + \left[ S_{n} - \ln p_{n} \right] D^{n}_{e} \, ,
	\label{eq:entBalanceEdge}
\end{equation}
which expresses the edge EP, the lhs, as the entropy change in each reservoir $r$ plus the system entropy change, the rhs.
See \S~\ref{ex:intensiveExtensive} for explicit examples of $\excMtx$ and $\st{f_{y}}$.

In the most general formulation, the internal entropy $S$, the conserved quantities $\st{\sq}$ (hence $\st{\excMtx^{y}_{e}}$), and their conjugated fields $\st{f_{y}}$, change in time.
Physically, this modeling corresponds to two possible ways of controlling a system:
either through $\st{\sq}$ or $S$ which characterize the system states,
or through $\st{f_{y}}$ which characterize the properties of the reservoirs.
Throughout the paper, we use the word ``driving'' to describe any of these time-dependent controls, while we refer to those systems that are not time-dependently driven as \emph{autonomous}.

\subsection{Network-Specific Conserved Quantities}

We now specify the procedure to identify the complete set of conserved quantities of a system.
In doing so, we extend the results of Ref.~\cite{polettini16}.
For this purpose, let $\st{\bm C_{\alpha}}$ for $\alpha = 1,\dots,\nof{\upalpha}$, be an independent set of network \emph{cycles}.
Algebraically, $\st{\bm C_{\alpha}}$ is a maximal set of independent vectors in $\ker D$,
\begin{equation}
	D^{n}_{e} \, C^{e}_{\alpha} = 0 \, , \quad \text{for all } \, n \, ,
	\label{eq:cyclesMatrix}
\end{equation}
in which at most one entry in each forward--backward transition pair is nonzero.
Since $D$ is $\st{-1,0,1}$-valued, $\st{\bm C_{\alpha}}$ can always be chosen in such a way that their entries are $\st{0,1}$.
In this representation, their $1$-entries identify sets of transitions forming loops.
In the examples, we will represent cycles using the set of forward transitions only, and negative entries denote transitions along the backward direction.
We denote the matrix whose columns are $\st{\bm C_{\alpha}}$ by $C \equiv \st{C^{e}_{\alpha}}$.

By multiplying the matrices $\excMtx$ and $C$, we obtain the $M$-matrix \cite{polettini16}:
\begin{equation}
	M^{y}_{\alpha} := \excMtx^{y}_{e} \, C^{e}_{\alpha} \, .
	\label{eq:M}
\end{equation}
This fundamental matrix encodes the \emph{physical topology} of the system.
It describes the ways in which the conserved quantities $\st{\sq}$ are exchanged between the reservoirs across the system, as its entries quantify the influx of $\st{y}$ along each cycle, $\alpha$.
The physical topology is clearly build on top of the network topology encoded in $C$. 

The basis vectors of the $\coker M$, are defined as the system conservation laws.
They are denoted by $\st{\bm \ell^{\lambda}}$ for $\lambda=1,\dots,\nof{\uplambda}$ where $\nof{\uplambda} := \dim \coker M$ and satisfy
\begin{equation}
	\ell^{\lambda}_{y} \, \excMtx^{y}_{e} \, C^{e}_{\alpha}
	= \ell^{\lambda}_{y} \, M^{y}_{\alpha}
	= 0 \, , \quad \text{for all } \, \alpha \, .
	\label{eq:conservationLaws}
\end{equation}
From (\ref{eq:cyclesMatrix}), this implies that $\bm \ell^{\lambda} \excMtx \in (\ker D)^{\perp}$.
Since $(\ker D)^{\perp} \equiv \coim D$, one can introduce a set of states-space vectors $\st{\bm L^{\lambda}}$---\emph{i.e.} state variables in the states space---which are mapped into $\st{\bm \ell^{\lambda} \excMtx}$ by the transpose of $D$:
\begin{equation}
	L^{\lambda}_{n} \, D^{n}_{e} = \ell^{\lambda}_{y} \, \excMtx^{y}_{e} \equiv {\textstyle\sum_{r}} \left\{ {\textstyle\sum_{\sqLabel}}\ell^{\lambda}_{(\sqLabel,r)} \, \excMtx^{(\sqLabel,r)}_{e} \right\}\, .
	\label{eq:components}
\end{equation}
The properties of the incidence matrix guarantee that each $\bm L^{\lambda}$ is defined up to a reference value, see \emph{e.g.} Ref.~\cite{*[][{, \S~6.2.}] {knauer11}}.
We thus confirm that $\st{L^{\lambda}}$ are \emph{conserved quantities} since Eq.~\eqref{eq:components} are their balance equations:
the lhs identifies the change of $\st{L^{\lambda}}$ in the system, while the rhs expresses their change in the reservoirs.
The thermodynamic implications of shifting the reference values of $\st{L^{\lambda}}$ are discussed in \S~\ref{sec:tep}.

Importantly, the vector space spanned by the conserved quantities, $\st{L^{\lambda}}$, encompasses the system quantities $\st{\sq}$.
They correspond to $\ell^{\sqLabel}_{y} \equiv \ell^{\sqLabel}_{(\sqLabel',r)} = \delta^{\sqLabel}_{\sqLabel'}$, so that the balance equations \eqref{eq:balanceGlobal} are recovered.
The remaining conservation laws arise from the interplay between the \emph{specific} topology of the network, $C$, and its coupling with the reservoirs, $\excMtx$, and we will refer to them as \emph{nontrivial}.
Only for these, the row vector $\bm \ell$ may depend on time since $M$ is a function of time, see \S~\ref{ex:cls} and the application in \S~\ref{sec:QPC}.

Variations in time of the system quantities $\st{\sq}$ induce changes in the matrix $M$.
If these changes cause a modification of the size of its cokernel, \emph{i.e.} a change in the number of conserved quantities, we say that the physical topology was altered.
We emphasize that these changes are not caused by changes in the network topology since this latter remains unaltered. 
An example of physical topology transformation is given in \S~\ref{ex:cls} and in the application in \S~\ref{sec:REG}, while one of network topology is discussed in \S~\ref{sec:motor}.

\paragraph*{Remark}
The introduction of the conserved quantities is akin to that of scalar potentials for irrotational fields in continuous space.
Indeed, the vector $\bm \ell^{\lambda} \excMtx$ replaces the field, $D\transpose$ plays the role of the gradient operator, and $\bm L^{\lambda}$ becomes the potential.
The condition expressed by Eq.~\eqref{eq:conservationLaws} is that of irrotational fields, as it tells us that $\bm \ell^{\lambda} \excMtx$ vanishes along all loops.

\subsection{Network-Specific Local Detailed Balance}
\label{sec:ldb}

We now make use of the conserved quantities, $\st{L^{\lambda}}$, to separate the conservative contributions in the local detailed balance \eqref{eq:ldb} from the nonconservative ones.
This central result will provide the basis for our EP decomposition in \S~\ref{sec:tep}.

We start by splitting the set $\st{y}$ into two groups:
a ``potential'' one $\st{\rf}$, and a ``force'' one $\st{\push}$.
The first must be constructed with $\nof{\uplambda}$ elements such that the matrix whose entries are $\st{\ell^{\lambda}_{\rf}}$ is nonsingular.
We denote the entries of the inverse of the latter matrix by $\st{\overline{\ell}^{\rf}_{\lambda}}$.
Crucially, since the rank of the matrix whose rows are $\st{\bm \ell^{\lambda}}$ is $\nof{\uplambda}$, it is always possible to identify a set of $\st{\rf}$.
However, it may not be unique and different sets have different physical interpretations, see \S\S~\ref{ex:decQD} and \ref{ex:epAwesome} as well as the following sections.
The second group, $\st{\push}$, is constructed with the remaining $\nof{\mathrm{y}} - \nof{\uplambda}$ elements of $\st{y}$.

With the above prescription, we can write the entries $\st{\excMtx^{\rf}_{e}}$ as functions of $\st{\excMtx^{\push}_{e}}$ and $\st{L^{\lambda}_{n}}$ by inverting $\st{\ell^{\lambda}_{\rf}}$ in Eq.~\eqref{eq:components},
\begin{equation}
	\excMtx^{\rf}_{e} = \overline{\ell}^{\rf}_{\lambda} \, L^{\lambda}_{n} \, D^{n}_{e} - \overline{\ell}^{\rf}_{\lambda} \ell^{\lambda}_{\push} \excMtx^{\push}_{e} \, .
	\label{}
\end{equation}
The local detailed balance \eqref{eq:ldb} can thus be rewritten as
\begin{equation}
	\ln \frac{w_{e}}{w_{-e}}
		= \phi_{n} D^{n}_{e}
		+ \mathcal{F}_{\push} \excMtx^{\push}_{e} \, .
	\label{eq:ldbAwesome}
\end{equation}
The first contribution is conservative since it derives from the \emph{potential}
\begin{equation}
	\phi_{n} := S_{n} - F_{\lambda} L^{\lambda}_{n} \, ,
	\label{eq:potential}
\end{equation}
where
\begin{equation}
	F_{\lambda} := f_{\rf} \overline{\ell}^{\rf}_{\lambda}
	\label{eq:effectiveIntensive}
\end{equation}
is a linear combination of entropic intensive fields.
Since $\phi_{n}$ is the entropy of the state $n$ minus a linear combination of conserved quantities, it can be viewed as the Massieu potential of the state $n$.
[We recall that Massieu potentials are the thermodynamic potentials of the entropy representation, see \emph{e.g.} \cite[\S~5-4]{callen85}.]
In contrast, the nonconservative \emph{fundamental forces},
\begin{equation}
	\mathcal{F}_{\push} := f_{\rf} \overline{\ell}^{\rf}_{\lambda} \ell^{\lambda}_{\push} - f_{\push} \, ,
	\label{eq:fundForce}
\end{equation}
are caused by the presence of multiple reservoirs.
As we will show, they control the currents of system quantities through the system.
Importantly, ``fundamental'' must be understood as a property of the set of these forces, since they are independent and in minimal number.

\begin{figure}[t]
	\centering
	\includegraphics[width=.42\textwidth]{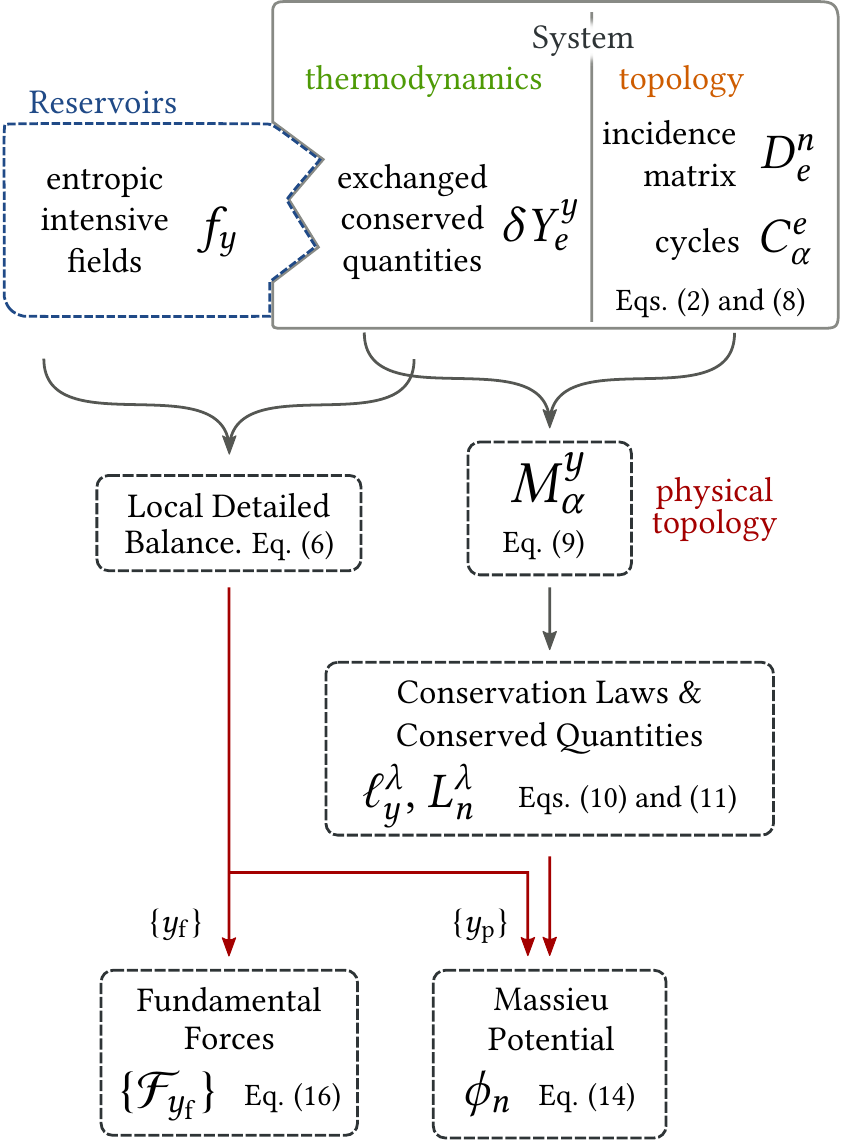}
	\caption{
		Schematic representation of our local detailed balance decomposition, which we summarize as follows.
		On the one hand, the system is characterized by those system quantities which are exchanged with the reservoirs along transitions, as well as by the topological properties of its network of transitions.
		The former is accounted for by the matrix of exchanged conserved quantities $\excMtx$, while the latter by the incidence matrix, $D$, Eq.~\eqref{eq:incidence}, which determines the matrix of cycles, $C$, Eq.~\eqref{eq:cyclesMatrix}.
		These two matrices combined give the $M$-matrix, Eq.~\eqref{eq:M}, which encodes the physical topology of the system and whose cokernel identifies the complete set of conservation laws and conserved quantities, Eq.~\eqref{eq:conservationLaws} and \eqref{eq:components}.
		On the other hand, the reservoirs are characterized by entropic intensive fields, $\st{f_{y}}$, which combined with the matrix of exchanged conserved quantities, $\excMtx$, gives the local detailed balanced, Eq.~\eqref{eq:ldb}.
		Having identified all conservation laws, the variables $y$ can be split into ``potential'' $y$, $\st{\rf}$, and ``force'' $y$, $\st{\push}$.
		The first group identifies a Massieu potential for each state $n$, $\phi_{n}$, Eq.~\eqref{eq:potential}, while the second one identifies the fundamental forces, Eq.~\eqref{eq:fundForce}.
		These two set of thermodynamic quantities are thus combined in the local detailed balanced, \eqref{eq:ldbAwesome}.
	}
	\label{fig:diagram}
\end{figure}

The identification of $\phi_{n}$ and $\st{\mathcal{F}_{\push}}$ and their relation with the local detailed balance, Eq.~\eqref{eq:ldbAwesome}, is the key result of our paper and we summarize the procedure we used in Fig.~\ref{fig:diagram}.
The complete set of conservation laws played an essential role in this identification.

We saw that driving in the system quantities $\st{\sq}$, may induce changes in the physical topology, whereas the driving in the reservoir properties, $\st{f_{y}}$,---as well as in the entropy, $S$---is unable to do so.
Since these changes modify the cokernel of $M$, $\phi_{n}$ and $\st{\mathcal{F}_{\push}}$ are modified as well:
when conservation laws are broken new fundamental forces emerge, and \emph{vice versa} the emergence of conservation laws breaks some fundamental forces and creates additional terms in $\phi_{n}$, see \S~\ref{ex:decQD}.

Even in absence of topological changes, the form of $\phi_{n}$ and $\st{\mathcal{F}_{\push}}$ may change in presence of driving.
It is clear that $\phi_{n}$ changes when $S$, $\st{\sq}$, or $\st{f_{\rf}}$ change, see Eq.~\eqref{eq:potential}.
In turn, each fundamental force $\mathcal{F}_{\push}$ depends on both $f_{\push}$ and $\st{f_{\rf}}$, see Eq.~\eqref{eq:fundForce}. 
But in presence of nontrivial conservation laws, they may also depend on the system quantities $\st{\sq}$ via the vectors $\st{\bm \ell^{\lambda}}$, see \S~\ref{ex:decQD} and the application in \S~\ref{sec:QPC}. 
Notice that while driving not caused by temperatures solely affects a given intensive field, driving via temperature, say $\beta_{r'}$, affects all the fields associated to $r'$, namely $\st{f_{(\sqLabel,r')}}$ for $\sqLabel=1,\cdots,\nof{\upkappa}$, see Tab.~\ref{tab:fYpairs}.

\subsection{Fundamental Cycles}
\label{sec:fundCycles}

We now express our conservative--nonconservative forces decomposition of the local detailed balance in terms of cycle affinities.
This provides the basis on which our potential--cycle affinities EP decomposition hinges on, \S~\ref{sec:tepSuper}.

The thermodynamic forces acting along cycles are referred to as \emph{cycle affinities}.
Using the local detailed balance \eqref{eq:ldbAwesome}, they read
\begin{equation}
	\mathcal{A}_{\alpha}
	:= C^{e}_{\alpha} \, \ln \frac{w_{e}}{w_{-e}}
	= \mathcal{F}_{\push} \, M^{\push}_{\alpha} \, .
	\label{eq:cycleAffinity}
\end{equation}
As observed in Ref.~\cite{polettini16}, different cycles may be connected to the same set of reservoirs, thus carrying the same cycle affinity.
These are regarded as \emph{symmetries} and they correspond to bases of $\ker M$, $\st{\bm \sym_{\rho}}$ for ${\rho = 1,\dots,\nof{\uprho} =: \dim \ker M}$,
\begin{equation}
	M^{y}_{\alpha} \, \sym^{\alpha}_{\rho} = 0 \, , \quad \text{for all } \, y \, ,
	\label{eq:symmetries}
\end{equation}
as their entries identify sets of cycles which, once completed, leave the state of the reservoirs unchanged.
A notable consequence is that the affinities corresponding to these sets of cycles are zero irrespective of the fields $\st{f_{y}}$.
The rank--nullity theorem applied to the matrix $M$ allows us to relate the number of symmetries to the number of conservation laws \cite{polettini16}
\begin{equation}
	\nof{\mathrm{y}} - \nof{\uplambda} = \nof{\upalpha} - \nof{\uprho} \, .
	\label{eq:rankNullity}
\end{equation}
Notice that, while the $\nof{\mathrm{y}}$ and $\nof{\upalpha}$ are fixed for a given system, $\nof{\uplambda}$, and hence $\nof{\uprho}$, can change due to changes in the physical topology.
From Eq.~\eqref{eq:rankNullity} we thus learn that for any broken (resp. created) conservation law, a symmetry must break (resp. be created), see \S~\ref{ex:cycles} and the application in \S~\ref{sec:REG}.

The symmetries given by Eq.~\eqref{eq:symmetries} lead us to identify $\nof{\upeta} := \nof{\upalpha} - \nof{\uprho}$ cycles, labeled by $\eta$, which correspond to linearly independent columns of $M$.
These cycles can be thought of as physically independent, since they cannot be combined to form cycles that leave the reservoirs unchanged upon completion.
In other words, they are the minimal subset of cycles whose affinity is nonzero for a generic choice of the fields $\st{f_{y}}$ (specific choices of $\st{f_{y}}$ can always make any cycle affinity equal to zero).
We refer to these cycles as \emph{fundamental cycles} and to their affinities as \emph{fundamental affinities}.
The fact that the matrix whose entries are $\st{M^{\push}_{\eta}}$ is square and nonsingular, see App.~\ref{sec:proofNonsingularity}, allows us to see the one-to-one correspondence between fundamental forces, Eq.~\eqref{eq:fundForce}, and these affinities,
\begin{equation}
	\mathcal{F}_{\push} = \mathcal{A}_{\eta} \overline{M}^{\eta}_{\push} \, ,
	\label{eq:F=-AM}
\end{equation}
where $\st{\overline{M}^{\eta}_{\push}}$ are the entries of the inverse matrix of that having $\st{M^{\push}_{\eta}}$ as entries.
In terms of $\st{\mathcal{A}_{\eta}}$, the local detailed balance, Eq.~\eqref{eq:ldbAwesome}, reads
\begin{equation}
	\ln \frac{w_{e}}{w_{-e}} 
	= \phi_{n} D^{n}_{e} + \mathcal{A}_{\eta} \zeta^{\eta}_{e} \, ,
	\label{eq:ldbSuperAwesome}
\end{equation}
where
\begin{equation}
	\zeta^{\eta}_{e} := \overline{M}^{\eta}_{\push} \, \excMtx^{\push}_{e}
	\label{eq:counterCycles}
\end{equation}
quantifies the contribution of each transition $e$ to the current along the fundamental cycle $\eta$ as well as all those cycles which are physically dependent on $\eta$.
Algebraically, the row vectors of $\zeta$, $\st{\bm \zeta^{\eta}}$, are dual to the physically independent cycles, $\st{\bm C_{\eta}}$,
\begin{equation}
	\zeta^{\eta}_{e} C^{e}_{\eta'} = \overline{M}^{\eta}_{\push} \, \excMtx^{\push}_{e} C^{e}_{\eta'} = \overline{M}^{\eta}_{\push} M^{\push}_{\eta'} = \delta^{\eta}_{\eta'}.
	\label{}
\end{equation}

Eq.~\eqref{eq:ldbSuperAwesome} is another key result of our paper, which expresses the conservative--nonconservative local detailed balance decomposition in terms of fundamental affinities.
Importantly, the affinities $\st{\mathcal{A}_{\eta}}$ depend on time both via $\st{f_{y}}$ and $\st{\sq}$, where the latter originates from the $M$-matrix, Eq.~\eqref{eq:cycleAffinity}.
Differently from $\st{\mathcal{F}_{\push}}$, they always have the dimension of an entropy.

\paragraph*{Remark}
Our set of fundamental cycles differs from that constructed with spanning trees and discussed by J.~Schnakenberg in Ref.~\cite{schnakenberg76}.
Algebraically, our set is not merely in $\ker D$, but rather in $\ker D \setminus \ker M$.
Furthermore, it is not constructed from the spanning trees of the graph.

\subsection{Detailed-Balanced Networks}
\label{sec:db}

We now focus on a specific class of dynamics called \emph{detailed balanced}.
These dynamics are such that either there are no forces ($\st{\push} = \varnothing$) or these are zero 
\begin{equation}
	\mathcal{F}_{\push} = f_{\rf} \bar{\ell}^{\rf}_{\lambda} \ell^{\lambda}_{\push} - f_{\push} = 0
	\label{eq:DB}
\end{equation}
---equivalently the affinities are zero, see Eq.~\eqref{eq:cycleAffinity}.
A driven detailed-balanced dynamics implies that the driving must keep the forces equal to zero at all times, while changing the potential $\phi_{n}$.
An autonomous detailed-balanced dynamics will always relax to an equilibrium distribution \cite{kolmogoroff36,kelly79}
\begin{equation}
	p^{\mathrm{eq}}_{n} = \exp\left\{ \phi_{n} - \Phi_{\mathrm{eq}} \right\} \, ,
	\label{eq:pEq}
\end{equation}
defined by the detailed balance property: $w_{e} p^{\mathrm{eq}}_{o(e)} = w_{-e} p^{\mathrm{eq}}_{o(-e)}$, for all $e$.
The last term, $\Phi_{\mathrm{eq}}$, is the logarithm of the partition function
\begin{equation}
	\Phi_{\mathrm{eq}} := \ln \left\{ {\textstyle\sum_{m}} \exp \left\{ \phi_{m} \right\} \right\} \, ,
	\label{eq:eqPotential}
\end{equation}
and can be identified as an \emph{equilibrium Massieu potential} \cite[\S\S~5-4 and 19-1]{callen85} \cite[\S~3.13]{peliti11}.

We now point out that one can transform a nondetailed-balance dynamics with the potential $\phi_{n}$ into a detailed-balanced dynamics with the same potential, if one can turn off the forces---set them to zero---without changing the potential.
This is is always possible through an appropriate choice of the fields $\st{f_{\push}}$, \emph{viz.} $f_{\push} = f_{\rf} \bar{\ell}^{\rf}_{\lambda} \ell^{\lambda}_{\push}$, except for the following cases:
when there are $f_{\push}$ such that $f_{\push} = \beta_{r'}$ (\emph{i.e.} $f_{\push}$ is the field conjugated with the exchange of energy with the reservoir $r'$) and $r'$ is among the reservoirs involved in $\st{\rf}$, then turning off the corresponding force $\mathcal{F}_{\push}$ via $f_{\push}$ will modify $\st{f_{\rf}}$ and in turn $\phi_{n}$.
Due to their importance for our FT, \S~\ref{sec:ftDFR}, we label these fields by $\st{{\push'}}$, to discriminate them from the other ones, denoted by $\st{{\push''}}$.
We finally observe that for isothermal processes all thermal gradients vanish beforehand, and one realizes that $\mathcal{F}_{\push'} = 0$ for all $\push'$, see \emph{e.g.} \S\S~\ref{sec:motor} and \ref{sec:REG}.
Therefore, turning off the forces never changes the potential.

\paragraph*{Remark}
The equilibrium distribution, Eq.~\eqref{eq:pEq}, is clearly the same one would obtain using a Maximum Entropy approach \cite{jaynes57} \cite[\S~3.17]{peliti11}.
Indeed, the distribution maximizing the entropy functional constrained by given values of the average conserved quantities $\st{\avef{L^{\lambda}} = L^{\lambda}}$,
\begin{multline}
	\mathcal{S}[p] = {\textstyle\sum_{n}} p_{n} \left[ S_{n} - \ln p_{n} \right] \\ - a \left( {\textstyle\sum_{n}} p_{n} - 1 \right) - a_{\lambda} \left( {\textstyle\sum_{n}} p_{n} L^{\lambda}_{n} - L^{\lambda} \right) \, ,
	\label{eq:PsiFunctional}
\end{multline}
is given by 
\begin{equation}
	p^{\ast}_{n} = \exp\left\{ S_{n} - a_{\lambda} L^{\lambda}_{n} - a \right\} \, .
	\label{}
\end{equation}
This is the equilibrium distribution, Eq.~\eqref{eq:pEq}, when the Lagrange multipliers are given by $a = \Phi_{\mathrm{eq}}$ and $a_{\lambda} = F_{\lambda}$, see Eq.~\eqref{eq:potential} and \eqref{eq:eqPotential}.

\section{Trajectory Level Description}
\label{sec:tep}

We now bring our description from the level of edges to trajectories.
A stochastic \emph{trajectory} of duration $t$, $\trj$, is defined as a set of transitions $\{e_{i}\}$ sequentially occurring at times $\{t_{i}\}$ starting from $n_{0}$ at time $0$.
If not otherwise stated, the transitions index $i$ runs from $i=1$ to the last transition prior to time $t$, $\nof{t}$, whereas the state at time $\tau \in [0,t]$ is denoted by $n_{\tau}$.
The values of $S$, $\st{\sq}$, and $\st{f_{y}}$ between time $0$ and an arbitrary time $t$ are all encoded in the \emph{protocol} $\pi_{\tau}$, for $\tau \in [0, t]$.

We first derive the balance for the conserved quantities, Eq.~\eqref{eq:components}.
The conservative and nonconservative contributions identified at the level of single transitions via the local detailed balance, Eqs.~\eqref{eq:ldbAwesome} and \eqref{eq:ldbSuperAwesome}, are then used to decompose the trajectory EP into its three fundamental contributions.

\subsection{Balance of Conserved Quantities}

Since the conserved quantities are state variables their change along a trajectory for a given protocol reads
\begin{equation}
	\begin{split}
		\Delta L^{\lambda}&[\trj] = L^{\lambda}_{n_{t}}(t) - L^{\lambda}_{n_{0}}(0) \\
		& = \int_{0}^{t} \de \tau \left\{ \at{\partial_{\tau} L^{\lambda}_{n}(\tau)}{n=n_{\tau}} + L^{\lambda}_{n}(\tau) D^{n}_{e} J^{e}(\tau) \right\} \, .
	\end{split}
	\label{}
\end{equation}
The first term on the rhs accounts for the instantaneous changes due to the time-dependent driving, while the second accounts for the finite changes due to stochastic transitions, since
\begin{equation}
	J^{e}(\tau) := {\textstyle\sum_{i}} \delta^{e}_{e_{i}} \delta(\tau - t_{i})
	\label{eq:instantaneousCurrent}
\end{equation}
are the trajectory-dependent instantaneous currents at time $\tau$.
Using the edge-wise balance, Eq.~\eqref{eq:components}, we can recast the above equation into
\begin{equation}
	\hspace{-1.2em}
		\Delta L^{\lambda}[\trj] = \int_{0}^{t} \de \tau \left\{ \at{\partial_{\tau} L^{\lambda}_{n}(\tau)}{n=n_{\tau}} + \ell^{\lambda}_{y}(\tau) \, I^{y}(\tau) \right\} \, ,
	\label{eq:trajBalance}
\end{equation}
where the physical currents
\begin{equation}
	I^{y}(\tau) := \excMtx^{y}_{e}(\tau) \, J^{e}(\tau) \, ,
	\label{eq:instPhenoCurrent}
\end{equation}
quantify the instantaneous influx of $y$ at time $t$.

\subsection{Entropy Balance}

The trajectory entropy balance is given by
\begin{align}
	&\Sigma[\trj]
	= \int_{0}^{t} \de \tau \, J^{e}(\tau) \ln \frac{w_{e}(\tau)}{w_{-e}(\tau)} - \ln \frac{p_{n_{t}}(t)}{p_{n_{0}}(0)} \label{eq:ep} \\
	&\hspace{-2em} = - \int_{0}^{t} \de \tau \, f_{y}(\tau) \excMtx^{y}_{e}(\tau) J^{e}(\tau) + \left[ \left( S_{n_{t}} - S_{n_{0}} \right) - \ln \frac{p_{n_{t}}(t)}{p_{n_{0}}(0)} \right] \, , \notag
\end{align}
As for the edge-wise balance, Eq.~\eqref{eq:entBalanceEdge}, the lhs is the EP, while the first and second term on the rhs are the entropy change of the reservoirs and the entropy change of the system \cite{schnakenberg76,seifert05}.
Using our decomposition of the local detailed balance, Eq.~\eqref{eq:ldbAwesome}, we can recast the latter equality into
\begin{align}
	\Sigma[\trj] & =
	- \ln \frac{p_{n_{t}}(t)}{p_{n_{0}}(0)} \label{eq:epIntermediate} \\
	& + \int_{0}^{t} \de \tau \left\{ \phi_{n}(\tau) D^{n}_{e} \, J^{e}(\tau) + \mathcal{F}_{\push}(\tau) \, I^{\push}(\tau) \right\} \, . \notag
\end{align}
Since $\phi_{n}$ is a state variable, its variations along the trajectory can be written as
\begin{multline}
	\Delta \phi[\trj] = \phi_{n_{t}}(t) - \phi_{n_{0}}(0) \\
	=  \int_{0}^{t} \de \tau \left\{ \phi_{n}(\tau) D^{n}_{e} J^{e}(\tau) + \at{ \partial_{\tau} \phi_{n}(\tau) }{n=n_{\tau}} \right\} \, .
	\label{eq:deltaphi}
\end{multline}
By combining Eqs.~\eqref{eq:epIntermediate} and \eqref{eq:deltaphi}, we can recast the trajectory EP in
\begin{equation}
	\Sigma[\trj] = 
	v[\trj]
	+ \Delta \Phi[\trj]
	+ {\textstyle\sum_{\push}} \sigma_{\push}[\trj] \, ,
	\label{eq:epAwesome}
\end{equation}
where
\begin{align}
	v[\trj]
		&:= - \int_{0}^{t} \de \tau \at{ \partial_{\tau} \phi_{n}(\tau) }{n=n_{\tau}} \, , \label{eq:Work} \\
	\Delta \Phi[\trj]
	&\phantom{:}= \Phi_{n_{t}}(t) - \Phi_{n_{0}}(0) \, , \\
	\sigma_{\push}[\trj]
	&:= \int_{0}^{t} \de \tau \, \mathcal{F}_{\push}(\tau) \, I_{\push}(\tau) \, ,  \label{eq:flowTypeDissRate} \, ,
\end{align}
with
\begin{equation}
	\Phi_{n} := \phi_{n} - \ln p_{n} \, .
	\label{eq:stochPotential}
\end{equation}

Eq.~\eqref{eq:epAwesome}, is the major result of our paper.
It shows the EP decomposed into a time-dependent driving contribution, a potential difference, and a minimal set of flux--force terms.
The first term only arises in presence of time-dependent driving.
It quantifies the entropy dissipated when $\phi_{n}$ is modified and we refer to it as the \emph{driving contribution}.
The second term is entirely conservative as it involves a difference between the final and initial \emph{stochastic Massieu potential}, Eq.~\eqref{eq:stochPotential}.
The last terms are nonconservative and prevent the systems from reaching equilibrium.
Each $\sigma_{\push}[\trj]$ quantifies the entropy produced by the flow of $\st{\push}$, and we refer to them as \emph{flow contributions}.

\begin{table}
	\centering
	\begin{tabular}{rccr}
		\toprule
		dynamics	& { }$v${ } 	& { }$\Delta \Phi${ } 	& { }$\sigma$ \\
		\midrule
		autonomous				 	& 0	&  		& 	\\
		NESS						& 0	& 0 	& 	\\
		driven detailed-balanced	& 	& 	 	& 0 \\
		autonomous detailed-balanced& 0	& 	 	& 0 \\
		\bottomrule
	\end{tabular}
	\caption{
		EP for common processes.
		``0'' denotes vanishing or negligible contribution, NESS is the acronym of \emph{nonequilibrium steady state}.
	}
	\label{tab:EPprocesses}
\end{table}

To develop more physical intuition of each single term, we now discuss them separately and consider some specific cases.
When writing the rate of driving contribution explicitly, Eq.~\eqref{eq:Work}, one obtains
\begin{equation}
	- \partial_{\tau} \phi_{n}
= - \partial_{\tau} S_{n} + \partial_{\tau} F_{\lambda} \, L^{\lambda}_{n} + F_{\lambda} \, \partial_{\tau} L^{\lambda}_{n} \, .
	\label{eq:workRate}
\end{equation}
When all $\st{\bm \ell^{\lambda}}$ are independent from system quantities, the terms, $\st{\partial_{\tau} F_{\lambda} \, L_{\lambda,n}}$, account for the entropy dissipated during the manipulation of the intensive fields $\st{f_{\rf}}$, Eq.~\eqref{eq:effectiveIntensive}.
In contrast, $\st{F_{\lambda} \, \partial_{\tau} L_{\lambda,n}}$ and $-\partial_{\tau} S_{n}$ characterize the dissipation due to the direct manipulation of the system quantities.
Clearly, the changes of those fields that do not appear in $\phi_{n}$ do not contribute to $v[\trj]$.

For autonomous processes, the EP becomes
\begin{equation}
	\Sigma[\trj] = 
	\Delta \Phi[\trj]
	+ \mathcal{F}_{\push} \mathcal{I}^{\push}[\trj] \, .
	\label{eq:epAwesomeNoDriving}
\end{equation}
where
\begin{equation}
	\mathcal{I}^{\push}[\trj] := \int_{0}^{t} \de \tau \, I^{\push} (\tau) \, ,
	\label{eq:intFundCurr}
\end{equation}
are the currents of $\st{\push}$ integrated along the trajectory.
The difference between the final and initial stochastic Massieu potential captures the dissipation due to changes of the internal state of the system.
For finite-dimensional autonomous processes, it is typically subextensive in time and negligible with respect to the nonconservative terms for long trajectories
\begin{equation}
	\Sigma[\trj] \overset{t\rightarrow \infty}{=}
	\mathcal{F}_{\push} \mathcal{I}^{\push}[\trj] \, .
	\label{eq:epAwesomeLongTime}
\end{equation}
The nonconservative flow contributions, Eqs.~\eqref{eq:flowTypeDissRate} and \eqref{eq:epAwesomeLongTime}, quantify the dissipation due to the flow of conserved quantities across the network.
Finally, for autonomous detailed-balanced systems, the nonconservative terms vanish, in agreement with the fact that these systems exhibit no net flows, and the EP becomes
\begin{equation}
	\Sigma[\trj] = 
	\Delta \Phi[\trj] \, .
	\label{eq:epAwesomeDB}
\end{equation}
Table~\ref{tab:EPprocesses} summarizes the contributions of the EP for these common processes.
We now proceed with three remarks.

\paragraph*{Remark}
We have already discussed the possibility of physical topology modifications due to driving, which consequently alter $\phi_{n}$ and $\st{\mathcal{F}_{\push}}$.
For protocols crossing points in which these modifications occur, the trajectory must be decomposed into subtrajectories characterized by the same physical-topology.
For each of these, our decomposition \eqref{eq:epAwesome} applies.

\paragraph*{Remark}
The contributions of the EP in Eq.~\eqref{eq:epAwesome} depend on the choice of $\st{\rf}$ and $\st{\push}$.
When aiming at quantifying the dissipation of a physical system, some choices may be more convenient than others depending on the experimental apparatus, see \emph{e.g.} \S~\ref{ex:epAwesome}.
This freedom can be thought of as a gauge of the EP.
In the long time limit, it only affects the flow contributions and it can be understood as a particular case of the gauge freedoms discussed in Refs.~\cite{polettini12,wachtel15}, which hinge on graph-theoretical arguments.

\paragraph*{Remark}
The driving contribution $v$ and the nonequilibrium Massieu potential $\Phi_{n}$ are defined up to a gauge.
This is evidenced when transforming the state variables $\st{L^{\lambda}}$ according to
\begin{equation}
	L^{\lambda}_{n}(t) \rightarrow U^{\lambda}_{\lambda'} \, L^{\lambda'}_{n}(t) + u^{\lambda} 1_{n} \, ,
	\label{}
\end{equation}
where $\st{U^{\lambda}_{\lambda'}}$ identify a nonsingular matrix, $\st{u^{\lambda}}$ are finite coefficients, and $\st{1_{n}}$ a vector whose entries are $1$.
The first term can be considered as a basis change of $\coker M$,
\begin{equation}
	\ell^{\lambda}_{y} \rightarrow U^{\lambda}_{\lambda'} \, \ell^{\lambda'}_{y} \, ,
	\label{eq:ellOmega}
\end{equation}
while the second as a \emph{reference shift} of $L^{\lambda}$.
Under the transformation \eqref{eq:ellOmega}, the fields \eqref{eq:effectiveIntensive} transform as
\begin{equation}
	F_{\lambda}(t) \rightarrow F_{\lambda'}(t) \, \overline{U}^{\lambda'}_{\lambda} \, ,
	\label{}
\end{equation}
where $U^{\lambda'}_{\lambda} \overline{U}^{\lambda}_{\lambda''} = \overline{U}^{\lambda'}_{\lambda} U^{\lambda}_{\lambda''} = \delta^{\lambda'}_{\lambda''}$, thus guaranteeing that scalar products are preserved.
As a consequence, the stochastic Massieu potential, Eq.~\eqref{eq:stochPotential}, and the rate of driving contribution, Eq.~\eqref{eq:workRate}, transform as
\begin{equation}
	\begin{aligned}
		\Phi_{n}(t) & \rightarrow \Phi_{n}(t) - \mathfrak{f}(t) 1_{n} \\
		- \partial_{t} \phi_{n}(t) & \rightarrow - \partial_{t} \phi_{n}(t) + \partial_{t} \mathfrak{f}(t) 1_{n} \, ,
	\end{aligned}
	\label{eq:gauge}
\end{equation}
where
\begin{equation}
	\mathfrak{f}(t) := F_{\lambda'}(t) \, \overline{U}^{\lambda'}_{\lambda} \, u^{\lambda} \, .
	\label{}
\end{equation}
Crucially, neither the local detailed balance \eqref{eq:ldbAwesome} nor the EP \eqref{eq:eprAwesome} are affected, as the physical process is not altered.
If only a basis change is considered, $\st{u^{\lambda} = 0}$, then $\mathfrak{f}(t) = 0$, and both $\Phi_{n}$ and $v$ are left unvaried.
Finally, for cyclic protocols, one readily sees that the driving work over a period is gauge invariant, since $\mathfrak{f}(t)$ is nonfluctuating.

The above gauge is akin to that affecting the potential--work connection and which led to several debates, see Ref.~\cite{campisi11} and references therein.
The problem is rooted in what is experimentally measured, as different experimental set-ups constrain to different gauge choices \cite{campisi11}.
We presented a general formulation of the gauge issue, by considering reference shifts of any conserved quantity, and not only of energy.

\subsection{Entropy Balance along Fundamental Cycles}
\label{sec:tepSuper}

An equivalent decomposition of the EP, Eq.~\eqref{eq:ep}, can be achieved using the potential--affinities decomposition of the local detailed balance, Eq.~\eqref{eq:ldbSuperAwesome}:
\begin{equation}
	\Sigma[\trj] = 
	v[\trj]
	+ \Delta \Phi[\trj]
	+ {\textstyle\sum_{\eta}} \gamma_{\eta}[\trj] \, .
	\label{eq:epSuperAwesome}
\end{equation}
Here,
\begin{equation}
	\gamma_{\eta}[\trj] := \int_{0}^{t} \de \tau \, \mathcal{A}_{\eta}(\tau) \, \zeta_{\eta,e} \, J^{e}(\tau) \, ,
	\label{eq:flowCycles}
\end{equation}
quantify the dissipation along the fundamental cycles, as $\st{\zeta_{\eta,e} \, J^{e}(\tau)}$, for ${\eta=1,\dots,\nof{\upeta}}$, are the corresponding instantaneous currents, Eq.~\eqref{eq:counterCycles}.
For autonomous processes, the EP becomes
\begin{equation}
	\Sigma[\trj] = 
	\Delta \Phi[\trj]
	+ \mathcal{A}_{\eta} \mathcal{Z}^{\eta}[\trj]
	\label{eq:epSuperAwesomeLT}
\end{equation}
where
\begin{equation}
	\mathcal{Z}^{\eta}[\trj] := \int_{0}^{t} \de \tau \, \zeta^{\eta}_{e} \, J^{e}(\tau)
	\label{eq:currFundCycle}
\end{equation}
measure the total circulation along $\st{\eta}$.

\section{Finite-time Detailed Fluctuation Theorem}
\label{sec:ftDFR}

The driving and flow contributions of the EP, Eq.~\eqref{eq:epAwesome}, are now shown to satisfy a finite-time detailed FT.
This constitutes another crucial result of our paper which generalizes previous FT formulations expressed in terms of physical currents.

\begin{figure}[t!]
	\centering
	\includegraphics[width=.48\textwidth]{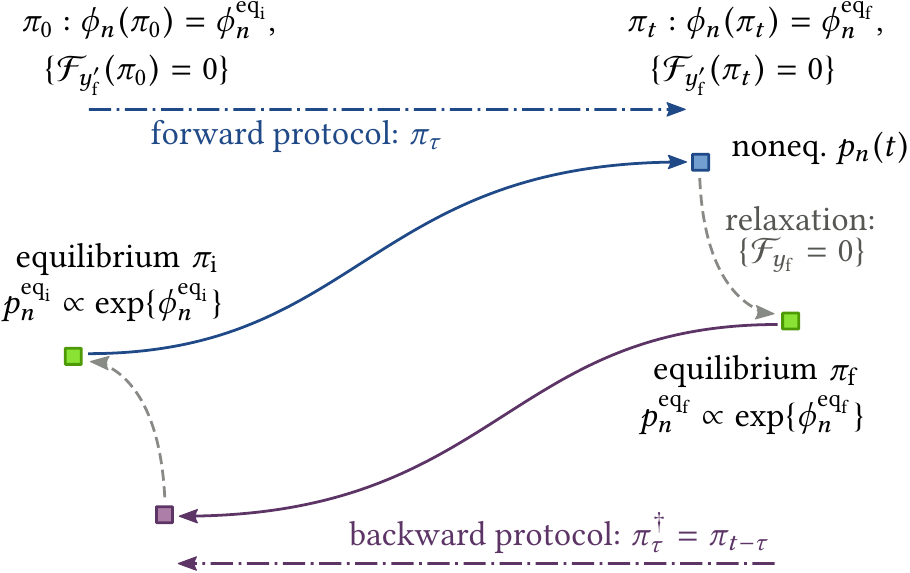}
	\caption{
		Schematic representation of the forward and backward processes.
	}
	\label{fig:fr}
\end{figure}

We consider a \emph{forward process} of duration $t$ defined as follows.
The system is initially prepared in an equilibrium state characterized by $\phi_{n}^{\mathrm{eq}_{\mathrm{i}}}$, Eq.~\eqref{eq:pEq}.
The latter state corresponds to the equilibrium protocol $\pi_{\mathrm{i}}$ in which $\phi_{n}(\pi_{\mathrm{i}}) = \phi_{n}^{\mathrm{eq}_{\mathrm{i}}}$ and naturally $\st{\mathcal{F}_{\push}(\pi_{\mathrm{i}}) = 0}$.
At time $\tau=0$ the protocol $\pi_{\tau}$, for $0 \le \tau \le t$, is activated.
It is arbitrary except at the boundaries, $\tau=0$ and $t$, where the following requirements must be satisfied:
at time $0$, the Massieu potential corresponding to $\pi_{0}$ must be the same as that of the initial equilibrium state, \emph{i.e.} $\phi_{n}(\pi_{0}) = \phi_{n}^{\mathrm{eq}_{\mathrm{i}}}$.
As a consequence, the fields $\st{f_{\push''}(\pi_{0})}$ can take arbitrarily values (i.e. they can be different from $\st{f_{\push''}(\pi_{\mathrm{i}})}$), while the other ones cannot: $\st{f_{\push'}(\pi_{0}) = f_{\push'}(\pi_{\mathrm{i}})}$.
This implies that $\st{\mathcal{F}_{\push''}(\pi_{0})}$ can be nonzero while $\st{\mathcal{F}_{\push'}(\pi_{0}) = 0}$.
Analogously, the protocol at time $t$ must be such that $\mathcal{F}_{\push'}(\pi_{t}) = 0$ for all $\push'$ while $\st{\mathcal{F}_{\push''}}$ can be arbitrary.
This condition guarantees that the Massieu potential $\phi_{n}(\pi_{t})$ identifies the equilibrium state corresponding to the equilibrium protocol $\pi_{\mathrm{f}}$: $\phi_{n}(\pi_{\mathrm{f}}) = \phi_{n}^{\mathrm{eq}_{\mathrm{f}}} = \phi_{n}(\pi_{t})$ and vanishing forces $\st{\mathcal{F}_{\push}(\pi_{\mathrm{f}}) = 0}$.
We can thus introduce the \emph{backward process} as that in which the system is initially prepared in the equilibrium state given by $\pi_{\mathrm{f}}$, and which is driven by the time-reversed protocol, $\pi^{\dagger}_{\tau} := \pi_{t - \tau}$, see Fig.~\ref{fig:fr}.

The \emph{finite-time detailed FT} states that the forward and backward process are related by
\begin{equation}
	\frac{P_{t}(v, \st{\sigma_{\push}})}{P^{\dagger}_{t}(-v, \st{-\sigma_{\push}})} = 
	\exp \left\{ v + {\textstyle\sum_{\push}} \sigma_{\push} + \Delta \Phi_{\mathrm{eq}} \right\} \, ,
	\label{eq:dftAwesome}
\end{equation}
where $P_{t}(v, \st{\sigma_{\push}})$ is the probability of observing a driving contribution of the EP equal to $v$, and flow ones $\st{\sigma_{\push}}$ along the forward process.
Instead, $P^{\dagger}_{t}(-v, \st{-\sigma_{\push}})$ is the probability of observing a driving contribution equal to $-v$, and flow ones $\st{-\sigma_{\push}}$ along the backward process.
The difference of equilibrium Massieu potentials, Eq.~\eqref{eq:eqPotential},
\begin{equation}
	\Delta \Phi_{\mathrm{eq}}
	= \Phi_{\mathrm{eq}_{\mathrm{f}}} - \Phi_{\mathrm{eq}_{\mathrm{i}}} \, ,
	\label{eq:deltaBetaPhiEq}
\end{equation}
refers to the final and initial equilibrium distributions.
When averaging over all possible values of $v$ and $\st{\sigma_{\push}}$, the integral FT ensues
\begin{equation}
	\ave{\exp \left\{ - v - {\textstyle\sum_{\push}} \sigma_{\push} \right\}} = \exp \left\{ \Delta \Phi_{\mathrm{eq}}  \right\}.
	\label{}
\end{equation}
We prove Eq.~\eqref{eq:dftAwesome} in App.~\ref{sec:proof} using a generating function technique which is new to our knowledge.

We now discuss insightful special cases of our general FTs.
We first consider those processes in which $\mathcal{F}_{\push'} = 0$ for all $\push'$ and at all times---isothermal processes are a notable instance---, the protocol can terminate without restrictions since $\phi_{n}(\pi_{\tau})$ always identifies an equilibrium state.
If, in addition, the protocol keeps the potential $\phi_{n}$ constant, \emph{viz.} $v = 0$, the FT \eqref{eq:dftAwesome} reads
\begin{equation}
	\frac{P_{t}(\st{\sigma_{\push}})}{P^{\dagger}_{t}(\st{-\sigma_{\push}})} = 
	\exp \left\{ {\textstyle\sum_{\push}} \sigma_{\push} \right\} \, .
	\label{eq:dftAwesomePhiConst}
\end{equation}
Yet a more detailed case is when the process is autonomous, for which we have
\begin{equation}
	\frac{P_{t}(\st{\mathcal{I}^{\push}})}{P_{t}(\st{-\mathcal{I}^{\push}})} = 
	\exp \left\{ \mathcal{F}_{\push} \mathcal{I}^{\push} \right\} \, ,
	\label{eq:dftAwesomeAutonomous}
\end{equation}
written in terms of integrated currents of $\st{\push}$, Eq.~\eqref{eq:intFundCurr}.
The latter FT can be seen as the result of having a constant protocol with nonvanishing the fundamental forces $\st{\mathcal{F}_{\push''}}$---but vanishing $\st{\mathcal{F}_{\push'}}$---operating on a system initially prepared at equilibrium.
Since nothing distinguishes the forward process from the backward one, the lhs is the ratio of the same probability distribution but at opposite values of $\st{\mathcal{I}^{\push}}$, see application in \S~\ref{sec:motor}.

Instead, for detailed-balanced systems we recover a Jarzynski--Crooks-like FT \cite{jarzynski97,crooks98} generalized to any form of time-dependent driving
\begin{equation}
	\frac{P_{t}(v)}{P^{\dagger}_{t}(-v)} = 
	\exp \left\{ v + \Delta \Phi_{\mathrm{eq}} \right\} \, .
	\label{eq:dftAwesomeDB}
\end{equation}

To provide a physical interpretation of the argument of the exponential on the rhs of Eq.~\eqref{eq:dftAwesome}, let us observe that once the protocol terminates, all fundamental forces can be switched off and the system relaxes to the equilibrium initial condition of the backward process. 
During the relaxation, neither $v$ nor $\st{\sigma_{\push}}$ evolve and the EP is equal to $\Phi_{\mathrm{eq}_{\mathrm{f}}} - \Phi_{n_{t}}$, Eq.~\eqref{eq:epAwesomeDB}.
Therefore, the argument of the exponential can be interpreted as the dissipation of the fictitious composite process ``forward process + relaxation to equilibrium''.

\paragraph*{Remark}
	As we discussed in Eq.~\eqref{eq:workRate}, the driving contribution consists of several subcontributions, one for each time-dependent parameter appearing in $\phi_{n}$.
	We formulated the finite-time FT \eqref{eq:dftAwesome} for the whole $v$, but it can be equivalently expressed for the single subcontributions, see \S~\ref{ex:ft}.

\subsubsection*{FT for Flow Contributions along Fundamental Cycles}
The FT \eqref{eq:dftAwesome} can also be expressed in terms of the flow contributions along the fundamental cycles $\st{\gamma_{\eta}}$ instead of $\st{\sigma_{\push}}$
\begin{equation}
	\frac{P_{t}(v, \st{\gamma_{\eta}})}{P^{\dagger}_{t}(-v, \st{-\gamma_{\eta}})} =
	\exp \left\{ v + {\textstyle\sum_{\eta}} \gamma_{\eta} + \Delta \Phi_{\mathrm{eq}} \right\} \, .
	\label{eq:dftSuperAwesome}
\end{equation}
Its proof is discussed in App.~\ref{sec:proof}.
The restrictions on $\pi_{0}$ and $\pi_{t}$ that we expressed in terms of $\st{\mathcal{F}_{\push}}$ can be re-expressed in terms of $\st{\mathcal{A}_{\eta}}$ via Eq.~\eqref{eq:F=-AM}.
For autonomous processes one can write the FT for the integrated currents along fundamental cycles, Eq.~\eqref{eq:currFundCycle},
\begin{equation}
	\frac{P_{t}(\st{\mathcal{Z}^{\eta}})}{P_{t}(\st{-\mathcal{Z}^{\eta}})} =
	\exp \left\{ \mathcal{A}_{\eta} \mathcal{Z}^{\eta} \right\} \, ,
	\label{eq:dftSuperAwesomeAuto}
\end{equation}
see Eq.~\eqref{eq:dftAwesomeAutonomous}.

\section{Ensemble Average Level Description}
\label{sec:EA}

We now discuss our results at the ensemble average level and derive a general formulation of the Nonequilibrium Landauer's Principle.

\subsection{Balance of Conserved Quantities}
 
Using the master equation \eqref{eq:ME} and the edge-wise balance \eqref{eq:components}, the balance equation for the average rates of changes of conserved quantities reads
\begin{equation}
	\dt \left[ {\textstyle\sum_{n}} L^{\lambda}_{n} \, p_{n} \right]
	\equiv \dt \avef{L^{\lambda}}
	= \avef{\dot{L}^{\lambda}} + \ell^{\lambda}_{y} \avef{I^{y}} \, ,
	\label{eq:averageBalance}
\end{equation}
where $\avef{\dot{L}^{\lambda}} := {\textstyle\sum_{n}} \partial_{t} L^{\lambda}_{n} \, p_{n}$ is the average change due to the driving, and
\begin{equation}
	\avef{I^{y}} := \excMtx^{y}_{e} \, \avef{J^{e}}
	\label{eq:phenoCurrent}
\end{equation}
are the average currents of $\st{y}$, see Eqs.~\eqref{eq:aveCurrents} and \eqref{eq:instPhenoCurrent}.
Hence, the second term in Eq.~\eqref{eq:averageBalance},
\begin{equation}
	\ell^{\lambda}_{y} \avef{I^{y}} = {\textstyle\sum_{r}} \left\{ {\textstyle\sum_{\sqLabel}} \ell^{\lambda}_{(\sqLabel,r)} \excMtx^{(\sqLabel,r)}_{e} \avef{J^{e}} \right\} \, ,
	\label{}
\end{equation}
accounts for the average flow of the conserved quantities in the reservoirs.
Obviously, the balances \eqref{eq:averageBalance} can also be obtained by averaging the trajectory balances \eqref{eq:trajBalance} along all stochastic trajectories.

\subsection{Entropy Balance}
In contrast to conserved quantities, entropy is not conserved.
The EP rate measures this nonconservation and is always non-negative
\begin{equation}
		\avef{\dot{\Sigma}}
		= {\textstyle\sum_{e}} w_{e} p_{o(e)} \ln \frac{w_{e} p_{o(e)}}{w_{-e} p_{o(-e)}} \ge 0 \, .
	\label{eq:epr}
\end{equation}
The EP decomposition in driving, conservative and flow contributions at the ensemble level, can be obtained by averaging Eq.~\eqref{eq:epAwesome}.
Alternatively, one can rewrite Eq.~\eqref{eq:epr} as
\begin{equation}
	\avef{\dot{\Sigma}}
	= - f_{y} \avef{I^{y}} + \left[ S_{n} - \ln p_{n} \right] D^{n}_{e} \avef{J^{e}} \, ,
	\label{eq:eprbis}
\end{equation}
where we used the local detailed balance property \eqref{eq:ldb} and the definition of average physical current \eqref{eq:phenoCurrent}.
The first term is the average entropy flow rate, while the second is the rate of change of the average system entropy.
Using the splitting of the set $\st{y}$ explained in \S~\ref{sec:st}, the physical currents of $\st{\rf}$ can be expressed as 
\begin{equation}
	\avef{I^{\rf}} = \overline{\ell}^{\rf}_{\lambda} \left[ \dt \avef{L^{\lambda}} - \avef{\dot{L}^{\lambda}} - \ell^{\lambda}_{\push} \avef{I^{\push}} \right] \, ,
	\label{}
\end{equation}
where we partially inverted Eq.~\eqref{eq:averageBalance}.
When combined with Eq.~\eqref{eq:eprbis}, the EP rate can be written as
\begin{equation}
	\avef{\dot{\Sigma}} 
	= \avef{\dot{v}}
	+ \dt \avef{\Phi}
	+ {\textstyle\sum_{\push}} \avef{\dot{\sigma}_{\push}}
	\, ,
	\label{eq:eprAwesome}
\end{equation}
where $\avef{\dot{v}} = - {\textstyle\sum_{n}} \partial_{t} \phi_{n} \, p_{n}$ is the driving contribution, $\avef{\dot{\sigma}_{\push}} = \mathcal{F}_{\push} \avef{I_{\push}}$ the flow contributions, and
\begin{equation}
	\avef{\Phi} = {\textstyle\sum_{n}} p_{n} \Phi_{n}
	\label{eq:NEMF}
\end{equation}
the \emph{nonequilibrium Massieu potential}.

Following a similar reasoning, and using the local detailed balance decomposition in terms of fundamental affinities, Eq.~\eqref{eq:ldbSuperAwesome}, we obtain the EP rate decomposed as
\begin{equation}
	\avef{\dot{\Sigma}} 
	= \avef{\dot{v}}
	+ \dt \avef{\Phi}
	+ {\textstyle\sum_{\eta}} \avef{\dot{\gamma}_{\eta}}
	\, ,
	\label{eq:eprSuperAwesome}
\end{equation}
where $\avef{\dot{\gamma}_{\eta}} = \mathcal{A}_{\eta} \zeta_{\eta,e} \avef{J^{e}}$ are the flow contributions along the fundamental cycles.

\subsection{Nonequilibrium Massieu Potential}

In detailed-balanced systems, the nonequilibrium Massieu potential takes its maximum value at equilibrium, Eq.~\eqref{eq:pEq}, where it becomes the equilibrium Massieu potential, Eq.~\eqref{eq:eqPotential}.
Indeed,
\begin{equation}
		\Phi_{\mathrm{eq}} - \avef{\Phi} = \avef{\Phi_{\mathrm{eq}} - \Phi} = \mathcal{D}(p\|p^{\mathrm{eq}}) \ge 0 \, ,
	\label{eq:relaxation}
\end{equation}
where
\begin{equation}
	\mathcal{D}(p\|p^{\mathrm{eq}}) := {\textstyle\sum_{n}} \, p_{n} \ln \frac{p_{n}}{p^{\mathrm{eq}}_{n}}
	\label{eq:relativeEntropy}
\end{equation}
is the relative entropy between the nonequilibrium distribution and the equilibrium one which quantifies the distance from equilibrium.

\paragraph*{Remark}
For autonomous detailed-balanced networks, the difference of equilibrium and nonequilibrium initial Massieu potential, Eq.~\eqref{eq:relaxation}, gives the average dissipation during the relaxation to equilibrium, $\avef{\Sigma} = \mathcal{D}(p(t_0)\|p_{\mathrm{eq}}) \ge 0$.
On the one hand, this shows how the MaxEnt principle mentioned in \S~\ref{sec:db} is embedded in the stochastic thermodynamic description (see also Ref.~\cite{altaner17}).
On the other hand, it underlines that its validity is limited to detailed-balanced systems.

\subsection{Nonequilibrium Landauer's Principle}
\label{sec:nelp}

\begin{figure}[t]
	\centering
	\includegraphics[width=.40\textwidth]{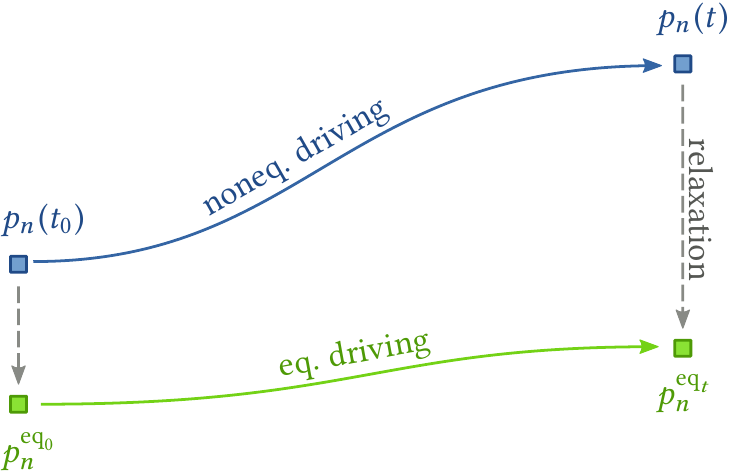}
	\caption{
		Schematic representation of the transformation between two nonequilibrium probability distributions.
		The protocol must leave the potential $\phi_{n}$ unchanged upon turning off of the forces at all times.
		This ensures that $\phi_{n}$ always identifies an equilibrium distribution (green curve) obtained by turning off the forces, shutting down the driving and letting the system relax (dashed gray curves).
		The nonequilibrium transformation---the blue curve---can be thus compared with the equilibrium one.
	}
	\label{fig:nelp}
\end{figure}

We now express Eq.~\eqref{eq:eprAwesome} in terms of a well defined equilibrium distribution, obtained by turning off the forces without modifying the potential $\phi_{n}$.
We already discussed that this procedure is always well defined for isothermal systems but requires more care for nonisothermal systems.
Combining Eqs.~\eqref{eq:eprAwesome} and \eqref{eq:relaxation}, one finds that 
\begin{equation}
	\avef{\dot{\Sigma}} 
	= \avef{\dot{v}_{\mathrm{irr}}}
	- \dt \mathcal{D}(p\|p^{\mathrm{eq}}) 
	+ {\textstyle\sum_{\push}} \avef{\dot{\sigma}_{\push}}
	\, ,
	\label{eq:eprAwesomeRefEq}
\end{equation}
where we introduced the average \emph{irreversible driving contribution} 
\begin{equation}
	\avef{\dot{v}_{\mathrm{irr}}} := \avef{\dot{v}} + \de_t \Phi_{\mathrm{eq}} \, .
	\label{eq:irrAveWork}
\end{equation}
Notice that the above contribution is not affected by the gauge discussed in \S~\ref{sec:tep}. 
Integrating Eq.~\eqref{eq:eprAwesomeRefEq} over time we get
\begin{equation}
	\avef{v_{\mathrm{irr}}} + {\textstyle\sum_{\push}} \avef{\sigma_{\push}}
	= \Delta \mathcal{D} ( p \| p_{\mathrm{eq}} ) + \avef{\Sigma} \, .
	\label{eq:NELPirr}
\end{equation}
This relation generalizes the nonequilibrium Landauer's principle, which is typically derived for driven detailed-balance systems, $\avef{\sigma_{\push}}=0$, \cite{esposito11}---see also Refs.~\cite{hasegawa10,takara10,altaner17}---, and which is used as the basis to study thermodynamics of information processing \cite{parrondo15}. 
It shows that not only driving but also flow EP must be consumed to move a system away from equilibrium, as depicted in Fig.~\ref{fig:nelp}, and that the minimal cost for doing so is precisely measured by the change in relative entropy. 
For driven detailed-balanced protocols connecting two equilibrium states, we recover the classical result that $\avef{\dot{v}_{\mathrm{irr}}} = \avef{\Sigma} \geq 0$.  

\subsection{Relation with previous EP decompositions}
\label{sec:relationPreviousEP}

We now briefly comment on the differences between our EP rate decomposition and other decompositions found in the literature. 

In Ref.~\cite{bulnescuetara14}, the obvious conserved quantities $\st{\sq}$ are used to express the EP rate in terms of a driving, a conservative, and a nonconservative term.
The first two are expressed in terms of a Massieu potential based on the $\nof{\upkappa}$ obvious conserved quantities, $\st{\sq}$, while the last is a sum of $\nof{\mathrm{y}} - \nof{\upkappa}$ flux--force contributions.
A finite-time FT solely expressed in terms of physical observable ensues.
In our work, by taking all $\nof{\uplambda}$ conserved quantities---trivial and nontrivial---into account, the nonconservative term is reduced to a sum of $\nof{\mathrm{y}} - \nof{\uplambda}$ fundamental flux--force contributions, and the new Massieu potential entering the driving and conservative contribution takes all conservation laws into account.
This has two crucial consequences for the ensuing FT:
\emph{(i)} our class of equilibrium distributions is broader since it is determined imposing a lower number of constraints, Eq.~\eqref{eq:DB} (\emph{i.e.} $\nof{\mathrm{y}} - \nof{\uplambda}$ vanishing forces instead of $\nof{\mathrm{y}} - \nof{\upkappa}$);
\emph{(ii)} the final value of the protocol must be constrained as discussed in \S~\ref{sec:ftDFR} since the new Massieu potential does not always identifies an equilibrium distribution.

In Ref.~\cite{polettini16} the Authors analyzed the reduction of flux--force contributions for systems at steady state, where the conservative contribution is absent.
Our decomposition \eqref{eq:eprAwesome} generalizes these results to nonautonomous systems in transient regimes.

In Refs.~\cite{schnakenberg76} and \cite{polettini14:cocycle}, decompositions based on graph-theoretic techniques are proposed, and the ensuing FTs are studied in Refs.~\cite{andrieux07:schnakenberg} and \cite{polettini14:transient}, respectively.
The nonconservative term of the EP rate is expressed as the sum of $\nof{\upalpha}$ cycle flux--affinity contributions.
These are typically in large number, see \emph{e.g.} \S\S~\ref{sec:motor} and \ref{sec:REG}.
Our decomposition \eqref{eq:eprSuperAwesome} demonstrates that only a subset of $\nof{\upalpha} - \nof{\uprho} = \nof{\mathrm{y}} - \nof{\uplambda}$ fundamental cycle flux--affinity contributions are necessary and sufficient to characterize the aforementioned term, where $\nof{\uprho}$ is the number of symmetries.

Yet a different EP decomposition is the \emph{adiabatic--nonadiabatic} one \cite{esposito07,harris07,esposito10:threedft,esposito10:threefaces1,ge10}.
Here, the driving and conservative terms arise from the stochastic potential $\Psi_{n} := - \ln \{ p_{n}/p^{\mathrm{ss}}_{n} \}$, which accounts for the mismatch between the actual and the steady-state probability distribution.
Instead, the nonconservative contribution quantifies the break of detailed balance of the steady state.
Hence, the steady-state probability distribution plays the role of a reference distribution in the same way that the equilibrium one (obtained by setting the forces to zero) does for our decomposition.
This is particularly clear when comparing \cite[Eq.~(21)]{esposito07} to Eq.~\eqref{eq:eprAwesomeRefEq}. Naturally, the equilibrium distribution is much more accessible than the steady-state one and implies that our decomposition is expressed in terms of physically measurable quantities.

\section{Applications}
\label{sec:applications}

We now analyze four model-systems:
a double quantum dot (QD), a QD coupled to a quantum point contact (QPC), a molecular motor, and a randomized grid.

\subsection{Double QD}
\label{sec:double}

This model has been extensively used in the past \cite{sanchez12,strasberg13,thierschmann15} and we will analyze it step by step following the order of the main text to illustrate of our formalism and our main results.

\begin{figure*}[t]
	\centering
	\subfloat[][Scheme]
	{\includegraphics[width=.30 \textwidth]{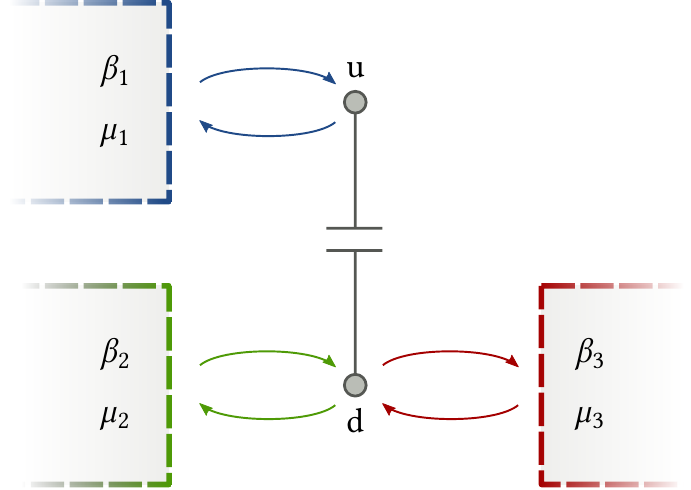} \label{fig:doubleDot} } \quad
	\subfloat[][Energy Landscape]
	{\includegraphics[width=.30\textwidth]{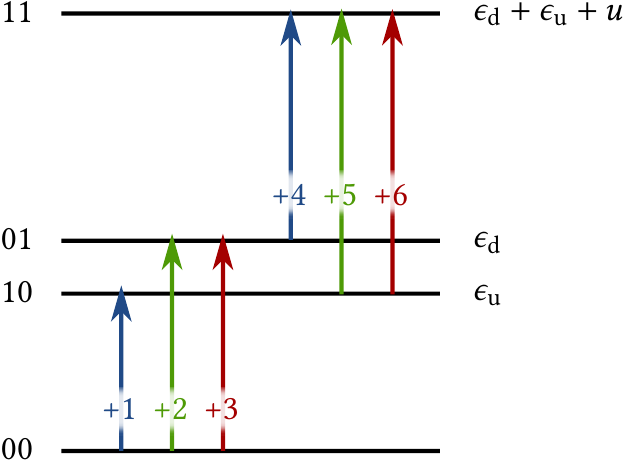} \label{fig:energy} } \quad
	\subfloat[][Transition Network]
	{\includegraphics[width=.30\textwidth]{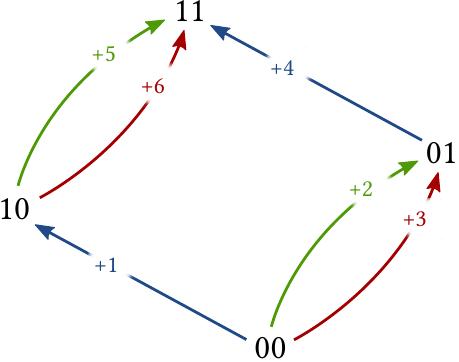} \label{fig:network} }
	\caption{
		Double QD coupled to three reservoirs and coupled with each other via a capacitor.
		Transitions related to the first reservoir are depicted in blue while those related to the second and third one by green and red, respectively.
		(a) Pictorial representation of the system.
		The upper dot $\mathrm{u}$ is coupled to the first reservoir, while the lower dot $\mathrm{d}$ is coupled to the second and third reservoir.
		The reservoirs exchange energy and electrons with the dots, which cannot host more than one electron.
		(b) Energy landscape of the dot.
		Importantly, when both dots are occupied, $11$, a repulsive energy $u$ adds to the occupied dots energies, $\epsilon_{\mathrm{u}}$ and $\epsilon_{\mathrm{d}}$.
		(c) Transition network of the model.
	}
	\label{fig:scheme}
\end{figure*}

\subsubsection{Setup}
\label{ex:intensiveExtensive}

The two single-level QDs is depicted in Fig.~\ref{fig:doubleDot}, whereas the energy landscape and the network of transitions are shown in Figs.~\ref{fig:energy} and \ref{fig:network}, respectively.
Electrons can enter empty dots from the reservoirs but cannot jump from one dot to the other.
When the two dots are occupied, an interaction energy, $u$, arises.

The network topology is encoded in the incidence matrix, whose representation in terms of the forward transitions reads
\begin{equation}
	D =
	\kbordermatrix{
	   & +1 & +2 & +3 & +4 & +5 & +6 \\
	00 & -1 & -1 & -1 & 0 & 0 & 0 \\
	10 & 1 & 0 & 0 & 0 & -1 & -1 \\
	01 & 0 & 1 & 1 & -1 & 0 & 0 \\
	11 & 0 & 0 & 0 & 1 & 1 & 1 
	} \, .
	\label{}
\end{equation}
Energy, $E_{n}$, and total number of electrons, $N_{n}$, characterize each system state:
\begin{equation}
	\begin{aligned}
		E_{00} & = 0 \, , & N_{00} & = 0 \, , \\
		E_{01} & = \epsilon_{\mathrm{d}} \, , & N_{01} & = 1 \, , \\
		E_{10} & = \epsilon_{\mathrm{u}} \, , & N_{10} & = 1 \, , \\
		E_{11} & = \epsilon_{\mathrm{u}} + \epsilon_{\mathrm{d}} + u \, , & N_{11} & = 2 \, ,
	\end{aligned}
	\label{eq:cqQD}
\end{equation}
where the first entry in $n$ refers to the occupancy of the upper dot while the second to the lower.
The entries of the matrix $\excMtx$ corresponding to the forward transitions are
\begin{equation}
	\excMtx =
	\kbordermatrix{
		& +1 & +2 & +3 & +4 & +5 & +6 \\
		(E,1) & \epsilon_{\mathrm{u}} & 0 & 0 & \epsilon_{\mathrm{u}} + u & 0 & 0 \\
		(N,1) & 1 & 0 & 0 & 1 & 0 & 0 \\
		(E,2) & 0 & \epsilon_{\mathrm{d}} & 0 & 0 & \epsilon_{\mathrm{d}} + u & 0 \\
		(N,2) & 0 & 1 & 0 & 0 & 1 & 0 \\
		(E,3) & 0 & 0 & \epsilon_{\mathrm{d}} & 0 & 0 & \epsilon_{\mathrm{d}} + u \\
		(N,3) & 0 & 0 & 1 & 0 & 0 & 1
	} \, ,
	\label{eq:exchangedQD}
\end{equation}
see Fig.~\ref{fig:network}, whereas the entries related to backward transition are equal to the negative of the forward.
For instance, along the first transition the system gains $\epsilon_{\mathrm{u}}$ energy and $1$ electron from the reservoir $1$.
The vector of entropic intensive fields is given by
\begin{equation}
	\bm {f} =
	\kbordermatrix{
		& (E,1) & (N,1) & (E,2) & (N,2) & (E,3) & (N,3) \\
		& \beta_{1} & - \beta_{1} \mu_{1} & \beta_{2} & - \beta_{2} \mu_{2} & \beta_{3} & - \beta_{3} \mu_{3}
	} \, .
	\label{eq:intensiveQD}
\end{equation}
Since the QDs and the electrons have no internal entropy, $S_{n} = 0$ for all $n$, the local detailed balance property, Eq.~\eqref{eq:ldb}, can be easily recovered from the product $- \bm f \excMtx$.
From a stochastic dynamics perspective, the latter property arises when considering fermionic transition rates:
$w_{e} = \Gamma_{e}(1+\exp\{ f_{y}\excMtx^{y}_{e} \})^{-1}$ and $w_{-e} = \Gamma_{e}\exp\{ f_{y}\excMtx^{y}_{e} \}(1+\exp\{ f_{y}\excMtx^{y}_{e} \})^{-1}$ for electrons entering and leaving the dot.

\subsubsection{Conservation Laws}
\label{ex:cls}

\begin{figure}[t]
	\centering
	\includegraphics[width=.47\textwidth]{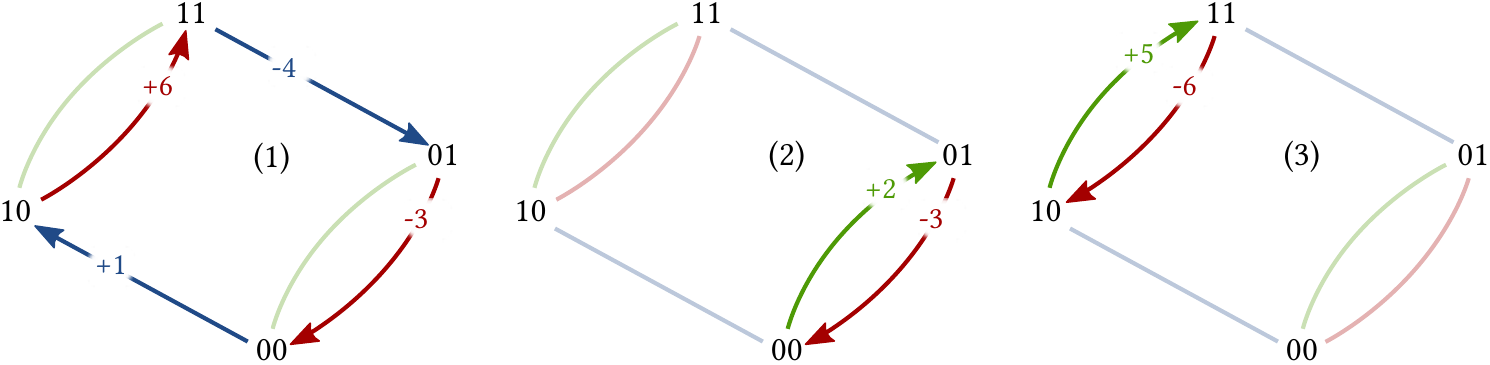}
	\caption{
		The independent set of cycles corresponding to the columns of $C$ in Eq.~\eqref{eq:cyclesQD}.
		The first corresponds to the sequence ``electron in $\mathrm{u}$ $\rightarrow$ electron in $\mathrm{d}$ $\rightarrow$ electron out of $\mathrm{u}$ $\rightarrow$ electron out of $\mathrm{d}$'', in which the lower QD is populated by the third reservoir.
		The second and third cycle correspond to the flow of one electron from the second reservoir to the third one, when the upper QD is empty and filled, respectively.
	}
	\label{fig:cyclesQD}
\end{figure}

We now illustrate the identification of the full set of conservation laws.
An independent set of cycles of this network, Fig.~\ref{fig:network}, is stacked in the matrix
\begin{equation}
	C =
	\kbordermatrix{
		& 1 & 2 & 3 \\
		+1 & 1 & 0 & 0 \\
		+2 & 0 & 1 & 0 \\
		+3 & -1 & -1 & 0 \\
		+4 & -1 & 0 & 0 \\
		+5 & 0 & 0 & 1 \\
		+6 & 1 & 0 & -1
	} \, ,
	\label{eq:cyclesQD}
\end{equation}
and corresponds to the cycles depicted in Fig.~\ref{fig:cyclesQD}.
The negative entries denote transitions performed in the backward direction.
The matrix encoding the physical topology, $M$, readily follows from the product of $\excMtx$ and $C$,
\begin{equation}
	M =
	\kbordermatrix{
		& 1 & 2 & 3 \\
		(E,1) & -u & 0 & 0 \\
		(N,1) & 0 & 0 & 0 \\
		(E,2) & 0 & \epsilon_{\mathrm{d}} & \epsilon_{\mathrm{d}} + u \\
		(N,2) & 0 & 1 & 1 \\
		(E,3) & u & -\epsilon_{\mathrm{d}} & -\epsilon_{\mathrm{d}}-u \\
		(N,3) & 0 & -1 & -1
	} \, .
	\label{eq:MQD}
\end{equation}
Its cokernel is spanned by
\begin{subequations}
	\begin{align}
		\bm \ell^{\mathrm{E}} & =
		\kbordermatrix{
			& (E,1) & (N,1) & (E,2) & (N,2) & (E,3) & (N,3) \\
			& 1 & 0 & 1 & 0 & 1 & 0
		} \label{eq:energyCL} \\
		\bm \ell^{\mathrm{u}} & =
		\kbordermatrix{
			& (E,1) & (N,1) & (E,2) & (N,2) & (E,3) & (N,3) \\
			& 0 & 1 & 0 & 0 & 0 & 0
		} \label{eq:upperCL} \\
		\bm \ell^{\mathrm{d}} & =
		\kbordermatrix{
			& (E,1) & (N,1) & (E,2) & (N,2) & (E,3) & (N,3) \\
			& 0 & 0 & 0 & 1 & 0 & 1
		} \, . \label{eq:lowerCL}
	\end{align}
	\label{eq:allCLQD}
\end{subequations}
The first vector identifies the energy state variable, $E_{n}$,
\begin{equation}
	\hspace{-1.2em}
	\bm \ell^{E} \excMtx = 
	\kbordermatrix{
		& +1 & +2 & +3 & +4 & +5 & +6 \\
		& \epsilon_{\mathrm{u}} & \epsilon_{\mathrm{d}} & \epsilon_{\mathrm{d}} & \epsilon_{\mathrm{u}} + u & \epsilon_{\mathrm{d}} + u & \epsilon_{\mathrm{d}} + u
	}
	\equiv \st{E_{n} D^{\mathrm{n}}_{e}} \, .
	\label{}
\end{equation}
The other two, instead, give the occupancy of the upper and lower dots, $N^{\mathrm{u}}_{n}$ and $N^{\mathrm{d}}_{n}$,
\begin{equation}
	\begin{aligned}
		\bm \ell^{\mathrm{u}} \excMtx & = 
		\kbordermatrix{
			& +1 & +2 & +3 & +4 & +5 & +6 \\
			& 1 & 0 & 0 & 1 & 0 & 0 \\
		}
		\equiv \st{N^{\mathrm{u}}_{n} D^{\mathrm{n}}_{e}} \, , \\
		\bm \ell^{\mathrm{d}} \excMtx & = 
		\kbordermatrix{
			& +1 & +2 & +3 & +4 & +5 & +6 \\
			& 0 & 1 & 1 & 0 & 1 & 1 \\
		}
		\equiv \st{N^{\mathrm{d}}_{n} D^{\mathrm{n}}_{e}} \, .
	\end{aligned}
	\label{}
\end{equation}
\emph{A posteriori}, we see that these conservation laws arise from the fact that no electron transfer from one dot to the other is allowed.
The total occupancy of the system, $N_{n}$, is recovered from the sum of the last two vectors.
Despite $\bm \ell^{\mathrm{u}}$ and $\bm \ell^{\mathrm{d}}$ are nontrivial conservation laws, they do not depend on any system quantity, Eq.~\eqref{eq:cqQD},
\footnote{
	One may argue that the above statement might be due the fact that we fixed the electron occupancy of each QD to one, Eq.~\eqref{eq:cqQD}.
	However, the same conclusion is reached when assuming:
	$N_{00} = 0$, $N_{01} = \nu_{\mathrm{d}}$, $N_{10} = \nu_{\mathrm{u}}$, and $N_{11} = \nu_{\mathrm{u}} + \nu_{\mathrm{d}}$, for some positive integer values $\nu_{\mathrm{u}}$ and $\nu_{\mathrm{d}}$.
}.

Let us now imagine that the interaction energy between the two dots is switched off, \emph{i.e.} $u\rightarrow 0$.
Two conservation laws emerge in addition to those in Eq.~\eqref{eq:allCLQD}:
\begin{subequations}
	\begin{align}
		\bm \ell^{(\mathrm{E},\mathrm{d})} & =
		\kbordermatrix{
			& (E,1) & (N,1) & (E,2) & (N,2) & (E,3) & (N,3) \\
			& 0 & 0 & 1 & 0 & 1 & 0
		} \label{eq:energyUCL} \\
		\bm \ell^{\mathrm{t}} & =
		\kbordermatrix{
			& (E,1) & (N,1) & (E,2) & (N,2) & (E,3) & (N,3) \\
			& 0 & 0 & -1 & \epsilon_{\mathrm{d}} & 0 & 0
		} \label{eq:tightCoupling}
		\, .
	\end{align}
	\label{eq:additionalCLQD}
\end{subequations}
The first is related to the upper--lower QD decoupling, as it corresponds to the conservation of energy of the lower dot
\begin{equation}
	\bm \ell^{(\mathrm{E},\mathrm{d})} \excMtx = 
	\kbordermatrix{
		& +1 & +2 & +3 & +4 & +5 & +6 \\
		& 0 & \epsilon_{\mathrm{d}} & \epsilon_{\mathrm{d}} & 0 & \epsilon_{\mathrm{d}} & \epsilon_{\mathrm{d}}
	}
	\equiv \st{E^{\mathrm{d}}_{n} D^{n}_{e}} \, .
	\label{}
\end{equation}
The conservation of energy in the upper dot is obtained as the difference between Eqs.~\eqref{eq:energyCL} and \eqref{eq:energyUCL}, and reads
\begin{equation}
	\bm \ell^{(\mathrm{E},\mathrm{u})} \excMtx = 
	\kbordermatrix{
		& +1 & +2 & +3 & +4 & +5 & +6 \\
		& \epsilon_{\mathrm{u}} & 0 & 0 & \epsilon_{\mathrm{u}} & 0 & 0
	}
	\equiv \st{E^{\mathrm{u}}_{n} D^{n}_{e}} \, .
	\label{}
\end{equation}
The second one, Eq.~\eqref{eq:tightCoupling}, arises from the tight coupling between the transport of energy and matter through the second dot.
Since $\bm \ell^{\mathrm{t}}$ is in $\coker \excMtx$,
\begin{equation}
	\bm \ell^{\mathrm{t}} \excMtx = 
	\kbordermatrix{
		& +1 & +2 & +3 & +4 & +5 & +6 \\
		& 0 & 0 & 0 & 0 & 0 & 0
	}
	\equiv \st{L^{\mathrm{t}}_{n} D^{n}_{e}} \, ,
	\label{}
\end{equation}
the conserved quantity $L^{\mathrm{t}}_{n}$ is a constant for all $n$, which can be chosen arbitrarily.
Notice the dependence on the system quantity $\epsilon_{\mathrm{d}}$ of the nontrivial conservation law \eqref{eq:tightCoupling}.
We thus showed that changes of system quantities ($u$ in our case) can modify the properties of $M$, and hence the set of conservation laws---without changing the network topology.

\subsubsection{Massieu Potential and Fundamental Forces}
\label{ex:decQD}

We now provide the expressions of $\phi_{n}$ and $\mathcal{F}_{\push}$ for the generic case $u\neq 0$.
Therefore, we split the set $\st{y}$ in $\st{\rf} = \st{(E,1), (N,1), (N,2)}$ and $\st{\push} = \st{(E,2), (E,3), (N,3)}$.
From Eq.~\eqref{eq:allCLQD} we see the validity of this splitting, as the matrix whose entries are $\st{\ell^{\lambda}_{\rf}}$ is an identity matrix.
The fields conjugated with the complete set of conservation laws, Eq.~\eqref{eq:effectiveIntensive}, are
\begin{align}
	F_{E} & = \beta_{1} \, , & F_{\mathrm{u}} & = - \beta_{1} \mu_{1} \, , & F_{\mathrm{d}} & = - \beta_{2} \mu_{2} \, ,
	\label{}
\end{align}
from which the Massieu potential of the state $n$, Eq.~\eqref{eq:potential}, follows
\begin{equation}
	\phi_{n} = - \beta_{1} E_{n} + \beta_{1} \mu_{1} N^{\mathrm{u}}_{n} + \beta_{2} \mu_{2} N^{\mathrm{d}}_{n} \, .
	\label{eq:potentialQD}
\end{equation}
Instead, the fundamental forces, Eq.~\eqref{eq:fundForce}, are given by
\begin{subequations}
	\begin{align}
		\mathcal{F}_{(E,2)} & = \beta_{1} - \beta_{2} \label{eq:effAffQD1} \, , \\
		\mathcal{F}_{(E,3)} & = \beta_{1} - \beta_{3} \label{eq:effAffQD2} \, , \\
		\mathcal{F}_{(N,3)} & = \beta_{3} \mu_{3} - \beta_{2} \mu_{2} \label{eq:effAffQD3} \, .
	\end{align}
	\label{eq:effAffQD}
\end{subequations}
The first two forces rule the energy flowing into the first reservoir from the second and third one, respectively, whereas the third force rules the electrons flowing from the third to the second reservoir.

Concerning the way the changes of $\phi_{n}$ and $\st{\mathcal{F}_{\push}}$ are intertwined, we see that the former depends on $\beta_{1}$, $\mu_{1}$, $\mu_{2}$, and $\beta_{2}$, which arises from $f_{(N,2)}$.
Therefore, while the changes of $f_{(E,3)} = \beta_{3}$ and $f_{(N,3)} = - \beta_{3} \mu_{3}$ only affect the related forces, the changes of $f_{(E,2)} = \beta_{2}$ affect both $\mathcal{F}_{(E,2)}$ and $\phi_{n}$.
Since the vectors of conservation laws \eqref{eq:allCLQD} do not depend on either $E_{n}$ or $N_{n}$, see \S~\ref{ex:cls}, the forces do not depend on system quantities.

Alternatively, one may split the set $\st{y}$ in $\st{\rf} = \st{(N,1), (E,2), (N,3)}$ and $\st{\push} = \st{(E,1), (N,2), (E,3)}$.
With this choice, we obtain
\begin{equation}
	\phi_{n} = - \beta_{2} E_{n} + \beta_{1} \mu_{1} N^{\mathrm{u}}_{n} + \beta_{3} \mu_{3} N^{\mathrm{d}}_{n} \, ,
	\label{eq:potentialQDsecond}
\end{equation}
and
\begin{subequations}
	\begin{align}
		\mathcal{F}_{(E,1)} & = \beta_{2} - \beta_{1} \, , \\
		\mathcal{F}_{(N,2)} & = \beta_{2} \mu_{2} - \beta_{3} \mu_{3} \, , \\
		\mathcal{F}_{(E,3)} & = \beta_{2} - \beta_{3} \, .
	\end{align}
	\label{eq:effAffQDdifferent}
\end{subequations}
With respect to the previous decomposition, we here consider the forces ruling the energy flow from the first and third reservoir, and the electrons flow from the second reservoir.

Let us now reconsider the case of vanishing interaction energy, $u=0$, as in \S~\ref{ex:cls}.
The five conservation laws that we consider are $E_{n}$, $E^{\mathrm{d}}_{n}$, $N^{\mathrm{u}}_{n}$, $N^{\mathrm{d}}_{n}$, $L^{\mathrm{t}}_{n}$, and we choose to split $\st{y}$ as $\st{\rf} = \st{(E,1),(N,1),(E,2),(N,2),(E,3)}$ and $\st{\push} = \st{(N,3)}$.
The potential follows
\begin{multline}
	\tilde{\phi}_{n} = - \beta_{1} E_{n} + \beta_{1} \mu_{1} N^{\mathrm{u}}_{n}
	+ \left[ \beta_{2} \mu_{2}  - \left( \beta_{2} - \beta_{3} \right) \epsilon_{\mathrm{d}} \right] N^{\mathrm{d}}_{n} \\
	- (\beta_{3} - \beta_{1}) E^{\mathrm{d}}_{n}
	- (\beta_{3} - \beta_{2}) L^{\mathrm{t}}_{n}
	\, ,
	\label{eq:phiQDu}
\end{multline}
whereas the only force is
\begin{equation}
	\tilde{\mathcal{F}}_{(N,3)} = \beta_{3} (\mu_{3} - \epsilon_{\mathrm{d}}) - \beta_{2} (\mu_{2} - \epsilon_{\mathrm{d}}) \, .
	\label{eq:forceQDu}
\end{equation}
We see that the creation of two conservation laws destroyed two nonconservative forces, Eqs.~\eqref{eq:effAffQD1} and \eqref{eq:effAffQD2}, whose expression can be spotted in the new potential, Eq.~\eqref{eq:phiQDu}.
Notice also how the emergence of the nontrivial conservation law \eqref{eq:tightCoupling} makes the fundamental force dependent on the system quantity $\epsilon_{\mathrm{d}}$.

\subsubsection{Symmetries and Fundamental Cycles}
\label{ex:cycles}

The two single-level QD has no symmetries for $u\neq 0$, since its $M$-matrix \eqref{eq:MQD} has empty kernel.
Its three cycle affinities, Eqs.~\eqref{eq:cyclesQD} and \eqref{eq:cycleAffinity}, are thus fundamental and read
\begin{subequations}
	\begin{align}
		\mathcal{A}_{1} & = \beta_{1} u - \beta_{3} u \, , \\
		\mathcal{A}_{2} & = \beta_{3} (\epsilon_{\mathrm{d}} - \mu_{3}) - \beta_{2} (\epsilon_{\mathrm{d}} - \mu_{2}) \label{eq:cycleAff2} \, , \\
		\mathcal{A}_{3} & = \beta_{3} (\epsilon_{\mathrm{d}} + u - \mu_{3}) - \beta_{2} (\epsilon_{\mathrm{d}} + u - \mu_{2}) \label{eq:cycleAff3} \, ,
	\end{align}
	\label{eq:cycleAff}
\end{subequations}
while the matrix relating fundamental cycles to edges, Eq.~\eqref{eq:counterCycles}, is given by
\begin{equation}
	\hspace{-1.2em}
	\zeta^{\eta}_{e} =
	\kbordermatrix{
		& +1 & +2 & +3 & +4 & +5 & +6 \\
		1 & 0 & \epsilon_{\mathrm{d}} & \epsilon_{\mathrm{d}} & 0 & \epsilon_{\mathrm{d}} + u & \epsilon_{\mathrm{d}} + u \\
		2 & 0 & - \epsilon_{\mathrm{d}} & - \epsilon_{\mathrm{d}} - u & 0 & - \epsilon_{\mathrm{d}} - u & - \epsilon_{\mathrm{d}} - u \\
		3 & 0 & \epsilon_{\mathrm{d}} & \epsilon_{\mathrm{d}} & 0 & \epsilon_{\mathrm{d}} + u & \epsilon_{\mathrm{d}}
	}
	\, \frac{1}{u} \, .
	\label{eq:monster}
\end{equation}
In sharp contrast with the fundamental forces, Eq.~\eqref{eq:effAffQD}, the fundamental affinities depend both on the fields and the system quantities.

As the interaction energy is turned off, two symmetries emerge:
\begin{subequations}
	\begin{align}
		\sym_{1} & =
		\kbordermatrix{
			& 1 & 2 & 3 \\
			& 1 & 0 & 0
		}
		\label{eq:symmetryQD1} \\
		\sym_{2} & =
		\kbordermatrix{
			& 1 & 2 & 3 \\
			& 0 & 1 & -1
		}
		\label{eq:symmetryQD2} \, ,
	\end{align}
\end{subequations}
in agreement with the creation of two conservation laws, see Eqs.~\eqref{eq:rankNullity} and \eqref{eq:additionalCLQD}.
They inform us that since the QDs are decoupled:
\emph{(i)} the cycle $1$ does not produces changes in the reservoirs, \emph{i.e.} its affinity is zero irrespective of the entries of $\bm f$;
\emph{(ii)} the cycle $2$ and $3$ are physically dependent since the flow of electrons from the second to the third reservoir is the same with empty and filled upper dot.
Choosing the third cycle as the fundamental one, its affinity reads as $\tilde{\mathcal{F}}_{(N,3)}$ in Eq.~\eqref{eq:forceQDu}, whereas the matrix of cycle contributions, see Eq.~\eqref{eq:counterCycles} and \S~\ref{ex:decQD}, becomes
\begin{equation}
	\zeta^{3}_{e} =
	\kbordermatrix{
		& +1 & +2 & +3 & +4 & +5 & +6 \\
		& 0 & 0 & -1 & 0 & 0 & -1 \\
	}
	\, .
	\label{eq:monsterU}
\end{equation}
Notice that both the transition $+3$---which belongs to the cycle $2$---and $+6$---which belongs to the cycle $3$---contribute to the current along the fundamental cycle $3$.

\subsubsection{Detailed-Balance Dynamics}
\label{ex:equilibrium}

From Eq.~\eqref{eq:effAffQD}, we see that the dynamics of the two QDs is detailed balanced when $\beta_{1} = \beta_{2} = \beta_{3}$ and $\mu_{2} = \mu_{3}$.
In this case the Massieu potential of state $n$, Eq.~\eqref{eq:potentialQD}, is given by
\begin{equation}
	\phi_{n} = - \beta_{1} \left( E_{n} - \mu_{1} N^{\mathrm{u}}_{n} - \mu_{2} N^{\mathrm{d}}_{n} \right) \, . 
	\label{eq:eqPotentialQD}
\end{equation}
The only element distinguishing the latter from that in Eq.~\eqref{eq:potentialQD} is the fact that $\beta_{2} = \beta_{1}$, which arises from $\mathcal{F}_{(E,2)} = 0$.
Therefore, a nondetailed-balanced dynamics described by the decomposition \eqref{eq:potentialQD}--\eqref{eq:effAffQD} can become detailed-balance without changing $\phi_{n}$ as long as $\mathcal{F}_{(E,2)} = 0$.
Instead, the decomposition in Eqs.~\eqref{eq:potentialQDsecond} and \eqref{eq:effAffQDdifferent} requires both $\mathcal{F}_{(E,1)}$ and $\mathcal{F}_{(E,3)}$ to be zero.

\subsubsection{EP decomposition}
\label{ex:epAwesome}

For the sake of illustrating our EP decomposition let us assume that only $E_{n}$, $\mu_{2}$, and $\beta_{3}$ change in time.
According to the expressions of $\phi_{n}$ and $\st{\mathcal{F}_{\push}}$ derived in \S~\ref{ex:decQD}, we can distinguish two driving contributions of the EP, Eqs.~\eqref{eq:Work} and \eqref{eq:workRate}:
\begin{equation}
	v[\trj] = v_{\mathrm{E}}[\trj] + v_{(N,2)}[\trj] \, ,
	\label{eq:sumWorkQD}
\end{equation}
where the first term,
\begin{equation}
	v_{\mathrm{E}}[\trj] := \beta_{1} \int_{0}^{t} \de \tau \at{\partial_{\tau} E_{n}(\tau)}{n_{\tau}} \, ,
	\label{eq:mechanicalWorkQD}
\end{equation}
is usually referred to as mechanical work in stochastic thermodynamics (up to $\beta_{1}$), while the second,
\begin{equation}
	v_{(N,2)}[\trj] := - \beta_{2} \int_{0}^{t} \de \tau \, \partial_{\tau} \mu_{2}(\tau) N^{\mathrm{d}}_{n_{\tau}} \, ,
	\label{eq:drivingWorkQD}
\end{equation}
is the entropy dissipated due to the change of the chemical potential of the second reservoir.
The flow contributions, Eq.~\eqref{eq:flowTypeDissRate}, are instead given by
\begin{subequations}
	\begin{align}
		\sigma_{(E,2)}[\trj] & = \mathcal{F}_{(E,2)} \int_{0}^{t} \de \tau \, I_{(E,2)}(\tau) \label{eq:flowQD1} \, , \\
		\sigma_{(E,3)}[\trj] & = \int_{0}^{t} \de \tau \, \mathcal{F}_{(E,3)}(\tau) I_{(E,3)}(\tau) \label{eq:flowQD2} \, , \\
		\sigma_{(N,3)}[\trj] & = \int_{0}^{t} \de \tau \, \mathcal{F}_{(N,3)}(\tau) I_{(N,3)}(\tau) \label{eq:flowQD3} \, ,
	\end{align}
	\label{eq:flowDissQD}
\end{subequations}
where, the forces are given in Eq.~\eqref{eq:effAffQD}, while the instantaneous currents of ${\push}$ are
\begin{subequations}
	\begin{align}
		I_{(E,2)} & = \epsilon_{\mathrm{d}} \left[ J^{+2} - J^{-2} \right] + (\epsilon_{\mathrm{d}} + u) \left[ J^{+5} - J^{-5} \right] \, , \\
		I_{(E,3)} & = \epsilon_{\mathrm{d}} \left[ J^{+3} - J^{-3} \right] + (\epsilon_{\mathrm{d}} + u) \left[ J^{+6} - J^{-6} \right] \, , \\
		I_{(N,3)} & = J^{+3} - J^{-3} + J^{+6} - J^{-6} \, .
	\end{align}
	\label{eq:fundCurrentsQD}
\end{subequations}
We thus see that the first and the second flow contribution, Eqs.~\eqref{eq:flowQD1} and \eqref{eq:flowQD2}, quantify the dissipation due to the energy flowing from the second and third reservoir to the first, respectively.
Analogously, the third contribution, Eq.~\eqref{eq:flowQD3}, characterizes the EP due to the flow of electrons from the third reservoir to the second.
The EP is thus the sum of the terms in Eqs.~\eqref{eq:sumWorkQD} and \eqref{eq:flowDissQD} plus a difference of stochastic Massieu potential, Eqs.~\eqref{eq:potentialQD} and \eqref{eq:stochPotential}.
We notice that the change in time of $\beta_{3}$ is accounted for by the second and third flows, Eqs.~\eqref{eq:flowQD2} and \eqref{eq:flowQD3}, while not by a driving contribution, as $\beta_{3}$ does not contribute to $\phi_{n}$, Eq.~\eqref{eq:potentialQD}.

It is worth noting that, from an experimental point of view, the driving contribution demands information on the states of the trajectory.
Instead, the flow contributions require the measurement of the energy flow in the second and third reservoir and the electron flow in the third.
Let us now compare the above decomposition with that based on a different choice of $\st{\rf,\push}$, \emph{e.g.} the second one made in \S~\ref{ex:decQD}.
In this case the driving contribution reads,
\begin{equation}
	v[\trj] = v_{\mathrm{E}}[\trj] + v_{(E,3)}[\trj]
	\label{}
\end{equation}
where
\begin{equation}
	v_{(E,3)}[\trj] := - \mu_{3} \int_{0}^{t} \de \tau \, \partial_{\tau} \beta_{3}(\tau) N^{\mathrm{d}}_{n_{\tau}} \, .
	\label{}
\end{equation}
The flow contributions read as in Eq.~\eqref{eq:flowDissQD} with forces given in Eq.~\eqref{eq:effAffQDdifferent} and other expressions for the currents.
Now, the measurement of the energy flow in the first and third reservoir, as well as the electron flow in the second reservoir, are required to quantify these terms in experiments.

To make the difference between the two choices even sharper, one can easily see that if the only quantity changing in time is $\mu_{2}$, the driving contribution of the second choice vanishes while that of the first does not.
Therefore, depending on the physical system and the experimental apparatus, one choice may be more convenient than another when it comes to estimating the dissipation.

\subsubsection{EP decomposition along Fundamental Cycles}
\label{ex:epSuperAwesome}

For the scenario described in the previous subsection, \S~\ref{ex:epAwesome}, the flow contributions along fundamental cycles \eqref{eq:flowCycles} read
\begin{subequations}
	\begin{align}
		\gamma_{1}[\trj] & = \int_{0}^{t} \de \tau \, \mathcal{A}_{1}(\tau) \zeta_{1,e} J^{e}(\tau) \label{eq:cycleflowQD1} \\
		\gamma_{2}[\trj] & = \int_{0}^{t} \de \tau \, \mathcal{A}_{2}(\tau) \zeta_{2,e} J^{e}(\tau) \label{eq:cycleflowQD2} \\
		\gamma_{3}[\trj] & = \int_{0}^{t} \de \tau \, \mathcal{A}_{3}(\tau) \zeta_{3,e} J^{e}(\tau) \label{eq:cycleflowQD3} \, ,
	\end{align}
	\label{eq:cycleFlowQD}
\end{subequations}
where the affinities are given in Eq.~\eqref{eq:cycleAff} and the cycle--edge coupling matrix $\zeta$ in Eq.~\eqref{eq:monster}.
Concerning their physical interpretation, the first contribution corresponds to the flow of energy from the third reservoir to the first, while the last two to the entropy dissipated when transferring electrons from the second reservoir to the third with empty and filled upper dot, respectively.

\subsubsection{Finite-time Detailed FT}
\label{ex:ft}

We now illustrate the conditions under which our FT applies to the coupled QDs.
The process must start from equilibrium, Eq.~\eqref{eq:pEq}:
all forces vanish and the potential is given in Eq.~\eqref{eq:eqPotentialQD}.
As the protocol is activated, it must leave the fields appearing in $\phi_{n}$, Eq.~\eqref{eq:potentialQD}, ($\beta_{1}$, $\beta_{2} \, ( = \beta_{1})$, $\mu_{1}$, and $\mu_{2}$) unchanged, but all the others can be set to arbitrary values.
Subsequently, all fields and system quantities controlled by $\pi_{\tau}$, for $0<\tau<t$, can change arbitrarily, until time $t$, in which the force in Eq.~\eqref{eq:effAffQD1} must be turned off.
This condition guarantees that the potential at time $t$ is of the form in Eq.~\eqref{eq:eqPotentialQD}, thus identifying a new equilibrium state.
When the above force vanishes at all times, one can formulate FTs like those in Eqs.~\eqref{eq:dftAwesomePhiConst} and \eqref{eq:dftAwesomeAutonomous}.

To simplify the application of the FT let us consider the conditions described in \S~\ref{ex:epAwesome}, with the further simplification that all temperatures are equal and constant:
only $E_{n}$ and $\mu_{2}$ change in time.
Since $\beta_{2} = \beta_{1}$ at all times, we do not need to worry about how the protocol terminates and the FT reads
\begin{multline}
	\frac{P_{t}(v_{\mathrm{E}}, v_{(N,2)}, \sigma_{(N,3)})}{P^{\dagger}_{t}(-v_{\mathrm{E}}, -v_{(N,2)}, -\sigma_{(N,3)})} \\
	= \exp\left\{ v_{\mathrm{E}} + v_{(N,2)} + \sigma_{(N,3)} + \Delta \Phi_{\mathrm{eq}} \right\} \, ,
\end{multline}
where the different contributions are given in Eqs.~\eqref{eq:mechanicalWorkQD}, \eqref{eq:drivingWorkQD}, and \eqref{eq:flowQD3}.
Notice that the contributions of $v$ appear separately in the above expression, but one can equivalently express the FT in terms of the full driving work $v$, Eq.~\eqref{eq:sumWorkQD}, as in the main discussion.

\subsubsection{FT for Flow Contributions along Fundamental Cycles}
\label{ex:FTcycles}

We saw in the previous example that the force $\mathcal{F}_{(E,2)}$, Eq.~\eqref{eq:effAffQD1}, must be zero at time $0$ and $t$ for the validity of the FT \eqref{eq:dftAwesome}, and at all times for the FTs \eqref{eq:dftAwesomePhiConst} and \eqref{eq:dftAwesomeAutonomous}.
Using Eq.~\eqref{eq:F=-AM} in combination with the inverse of the submatrix of \eqref{eq:MQD} whose entries are $\st{M^{\push}_{\eta}}$,
\begin{equation}
	\overline{M} = 
	\kbordermatrix{
		& (E,2) & (E,3) &(N,3) \\
		1 	& 1 & 1 & 0 \\
		2 	& -1 & 0 & - \epsilon_{\mathrm{d}} - u \\
		3 	& 1 & 0 & \epsilon_{\mathrm{d}} 
	} \frac{1}{u} \, ,
	\label{}
\end{equation}
we conclude that the above requirement becomes
\begin{equation}
	\mathcal{A}_{1} - \mathcal{A}_{2} + \mathcal{A}_{3} = 0 \, ,
	\label{}
\end{equation}
in terms of fundamental affinities, Eq.~\eqref{eq:cycleAff}.
Once identified the above condition, the application of the FT readily follows.

\subsection{QD coupled to a QPC}
\label{sec:QPC}

\begin{figure}[t]
	\centering
	\includegraphics[width=.31\textwidth]{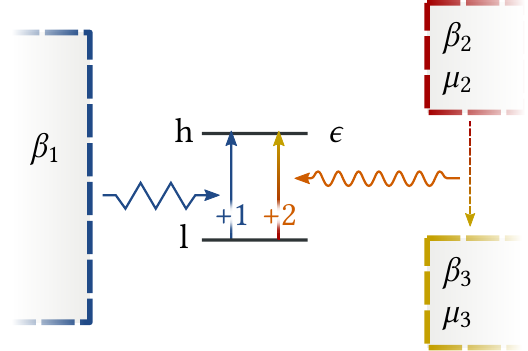}
	\caption{
		Model of QD coupled with a thermal reservoir and a pair of particle reservoirs modeling a QPC.
		The electron can jump to the excited state following either a phononic interaction with the first reservoir or an interaction with the QPC.
		The latter involves an electron current from the second to the third reservoir.
	}
	\label{fig:weird}
\end{figure}

We now consider a simplified description of a two levels QD coupled to a thermal reservoir and a QPC, Fig.~\ref{fig:weird}.
For a detailed analysis of this class of systems we refer to Ref.~\cite{bulnescuetara15}.
The interest of this model is twofold, it shows how single transitions can trigger exchanges involving multiple reservoir, and it also provides a further instance of a fundamental force which depends on system quantities due to nontrivial conservation laws.

The two states of the QD, $\mathrm{l}$ for ``low'' and $\mathrm{h}$ for ``high'', are characterized by different energies but the same number of electrons
\begin{equation}
	\begin{aligned}
		E_{\mathrm{l}} & = 0 \, , & E_{\mathrm{h}} & = \epsilon \, , & N_{\mathrm{l}} & = 1 \, , & N_{\mathrm{h}} & = 1 \, .
	\end{aligned}
	\label{eq:extenQuantQPC}
\end{equation}
The transition between these states can occur following either a phononic interaction with the first reservoir, $\pm1$, or following electron tunneling from the second to the third reservoir, $\pm2$.
Along the latter transition, an electron with energy $u + \epsilon$ leaves the second reservoir and enters the third with energy $u$.
The matrix of exchanged conserved quantities, $\excMtx$, thus reads
\begin{equation}
	\excMtx =
	\kbordermatrix{
		& +1 & +2 \\
		(E,1) & \epsilon & 0 \\
		(E,2) & 0 & u + \epsilon \\
		(N,2) & 0 & 1 \\
		(E,3) & 0 & -u \\
		(N,3) & 0 & -1
	} \, ,
	\label{eq:Yqpc}
\end{equation}
while the vector of intensive fields is
\begin{equation}
	\bm {f} =
	\kbordermatrix{
		& (E,1) & (E,2) & (N,2) & (E,3) & (N,3) \\
		& \beta_{1} & \beta_{2} & - \beta_{2} \mu_{2} & \beta_{3} & - \beta_{3} \mu_{3}
	} \, .
	\label{}
\end{equation}
The nontrivial local detailed balance property for the second transition follows from $ - \bm f \excMtx$, and reads
\begin{equation}
	\ln\frac{w_{+2}}{w_{-2}} = - \beta_{2} (u + \epsilon - \mu_{2}) + \beta_{3} (u - \mu_{3}) \, .
	\label{}
\end{equation}
The $M$-matrix,
\begin{equation}
	M =
	\kbordermatrix{
		& 1 \\
		(E,1) & \epsilon \\
		(E,2) & -u-\epsilon \\
		(N,2) & -1 \\
		(E,3) & u \\
		(N,3) & 1
	} \, ,
	\label{eq:Mqpc}
\end{equation}
follows from the product of $\excMtx$, Eq.~\eqref{eq:Yqpc}, and the matrix of cycles,
\begin{equation}
	C =
	\kbordermatrix{
		& 1 \\
		+1 & 1 \\
		+2 & -1 
	} \, .
	\label{eq:cyclesWeird}
\end{equation}
Its four-dimensional cokernel is spanned by
\begin{subequations}
	\begin{align}
		\bm \ell^{\mathrm{E}} & =
		\kbordermatrix{
			& (E,1) & (E,2) & (N,2) & (E,3) & (N,3) \\
			& 1 & 1 & 0 & 1 & 0
		} \, , \label{eq:energyCLqpc} \\
		\bm \ell^{\mathrm{N}} & =
		\kbordermatrix{
			& (E,1) & (E,2) & (N,2) & (E,3) & (N,3) \\
			& 0 & 0 & 1 & 0 & 1
		} \, , \label{eq:numberCLqpc} \\
		\bm \ell^{\mathrm{3}} & =
		\kbordermatrix{
			& (E,1) & (E,2) & (N,2) & (E,3) & (N,3) \\
			& 0 & 1 & - u - \epsilon & 0 & 0
		} \, , \label{eq:CLqpc1} \\
		\bm \ell^{\mathrm{4}} & =
		\kbordermatrix{
			& (E,1) & (E,2) & (N,2) & (E,3) & (N,3) \\
			& 0 & 0 & u & 1 & 0
		} \, . \label{eq:CLqpc2}
	\end{align}
	\label{}
\end{subequations}
The first two conservation laws are clearly the energy and the number of particles, Eq.~\eqref{eq:extenQuantQPC}, since $\bm \ell^{\mathrm{E}} \excMtx = (\epsilon,\epsilon)$ and $\bm \ell^{\mathrm{N}} \excMtx = (0,0)$.
For the other two, $\bm \ell^{3} \excMtx = \bm \ell^{4} \excMtx = (0,0)$ implies that the related conserved quantities are constants, \emph{i.e.} they do not depend on $n$.
Mindful of the gauge freedom described in \S~\ref{sec:tep} we can set the conserved quantities related to $\bm \ell^{\mathrm{N}}$, $\bm \ell^{3}$, and $\bm \ell^{4}$ to zero.
When $(E,1)$ is set as ``force'' $y$, the field related to the energy conservation law
\begin{equation}
	F_{\mathrm{E}} = \left[( \epsilon + u - \mu_2 )\beta_2 - ( u - \mu_3 )\beta_3 \right]/\epsilon \, ,
	\label{}
\end{equation}
determine the values of the nonequilibrium Massieu potential, $\phi_{n} = - F_{\mathrm{E}} E_{n}$.
Concerning the nonconservative contributions, the fundamental force and the fundamental affinity read
\begin{equation}
	\mathcal{F}_{(E,1)} = F_{\mathrm{E}} - \beta_{1} = \epsilon \mathcal{A}_{1} \, .
	\label{}
\end{equation}
Due to the emergence of nontrivial conservation laws, Eqs.~\eqref{eq:CLqpc1} and \eqref{eq:CLqpc2}, the fundamental force depends on a system quantity.
In detailed balance dynamics, $\mathcal{F}_{(E,1)} = 0$, and we readily recover $\phi_{n} = - \beta_{1} E_{n}$.

\subsection{Molecular Motor}
\label{sec:motor}

\begin{figure*}[t]
	\centering
	\includegraphics[width=\textwidth]{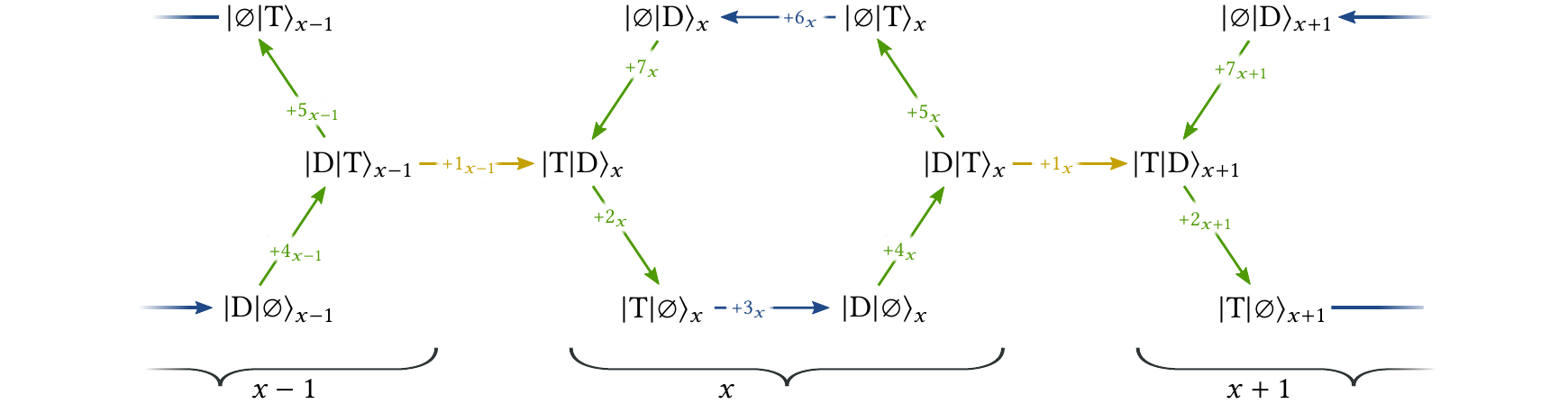}
	\caption{
		Network of transitions describing the chemomechanical stepping of the motor, where $x$ denotes the generic position along the stepping support.
		The molecular motor has six internal conformations distinguished by the state of the trailing, $|\cdot|$, and leading, $|\cdot\rangle$, motor foot:
		ATP-bound (T), ADP-bound (D), or unbound ($\varnothing$).
		Yellow arrows denote stepping transitions, $\st{+1_{x} \equiv |\ce{D}|\ce{T}\rangle_{x} \rightarrow |\ce{T}|\ce{D}\rangle_{x+1}}$, along which the mechanical force $k$ acts (positive value drive the system toward increasing $x$).
		Internal transitions may entail the exchange of $\ce{ATP}$ and $\ce{ADP}$ molecules with particle reservoirs (green arrows) or the hydrolysis of $\ce{ATP}$ into $\ce{ADP}$ (blue arrows).
		The latter only exchange energy with the thermal reservoir at inverse temperature $\beta$.
	}
	\label{fig:networkKinesin}
\end{figure*}

We now turn to the thermodynamic description of a molecular motor moving along a single dimension, see Refs.~\cite{liepelt07,altaner15}.
Beside providing an instance of a work reservoir, this model also illustrates 
how changes in the topology of the network can convert a conservative force into a nonconservative one.

The motor conformations and transitions are described in Fig.~\ref{fig:networkKinesin}.
It can step against a mechanical force $k$ thanks to the chemical force produced by the hydrolysis of $\ce{ATP}$ into $\ce{ADP}$, which are exchanged with reservoirs at chemical potential $\mu_{\ce{ATP}}$ and $\mu_{\ce{ADP}}$.
We label each state of the process by $n=(m,x)$, while each transition by $e_{x}$, where $e\in\st{1,2,3,4,5,6,7}$ refers to the transitions at a given position $x\in\mathbb{Z}$.
The system quantities are the internal energy, $E_{n} = \epsilon_{m}$, the total number of $\ce{ATP}$ plus $\ce{ADP}$ molecules attached to the motor, $N_{n} = N_{m}$, and the position, $X_{n} = xl$ where $l$ is the size of a step.
Importantly, each internal state is characterized by an internal entropy $S_{n} = s_{m}$.

The matrix of exchanged conserved quantities for the transitions at given position $x$ is written as
\begin{widetext}
	\begin{equation}
		\excMtx_{x} =
		\kbordermatrix{
			& +1_{x} & +2_{x} & +3_{x} & +4_{x} & +5_{x} & +6_{x} & +7_{x} \\
			(E) 			& \epsilon_{\mathrm{TD}} - \epsilon_{\mathrm{DT}} & \epsilon_{\mathrm{T}\varnothing}-\epsilon_{\mathrm{TD}} & \epsilon_{\mathrm{D}\varnothing}-\epsilon_{\mathrm{T}\varnothing} & \epsilon_{\mathrm{DT}}-\epsilon_{\mathrm{D}\varnothing} & \epsilon_{\varnothing\mathrm{T}}-\epsilon_{\mathrm{DT}} & \epsilon_{\varnothing\mathrm{D}}-\epsilon_{\varnothing\mathrm{T}} & \epsilon_{\mathrm{TD}}-\epsilon_{\varnothing\mathrm{D}} \\
			(N,\ce{ATP}) 	& 0 & 0 & 0 & 1 & 0 & 0 & 1 \\
			(N,\ce{ADP}) 	& 0 & -1 & 0 & 0 & -1 & 0 & 0 \\
			(X) 			& l & 0 & 0 & 0 & 0 & 0 & 0
		} \, ,
		\label{eq:Ykinesin}
	\end{equation}
\end{widetext}
whereas the full matrix is given by $\excMtx = \begin{pmatrix} \dots & \excMtx_{x-1} & \excMtx_{x} & \excMtx_{x+1} & \dots \end{pmatrix}$.
On the other side, the row vector of intensive variables reads
\begin{equation}
	\bm {f} =
	\kbordermatrix{
		& (E) & (N,\ce{ATP}) & (N,\ce{ADP}) & (X) \\
		& \beta & - \beta \mu_{\ce{ATP}} & - \beta \mu_{\ce{ADP}} & - \beta k
	} \, .
	\label{}
\end{equation}
Differently from all previous cases, the local detailed balance of the step transitions involves the \emph{work reservoir}, $(X,-\beta k)$,
\begin{equation}
	\ln \frac{w_{+1_{x}}}{w_{-1_{x}}} = - \beta \left[ \left( \epsilon_{\mathrm{TD}} - \epsilon_{\mathrm{DT}} \right) - k l \right] + \left( s_{\mathrm{TD}} - s_{\mathrm{DT}} \right) \, .
	\label{eq:ldbKinesin}
\end{equation}
Notice that the interpretation of the first term as minus entropy flow still holds:
$q_{+1_{x}} := \left( \epsilon_{\mathrm{TD}} - \epsilon_{\mathrm{DT}} \right) - k l$ is the heat of transition, since the last term is minus the work that the mechanical force exerts on the system \cite{seifert11,horowitz16}.

It is easily shown that the subnetwork at given $x$ contains exactly one cycle $c_{x}$,
\begin{equation}
	C_{x} =
	\kbordermatrix{
		& c_{x} \\
		+1_{x} & 0 \\
		+2_{x} & 1 \\
		+3_{x} & 1 \\
		+4_{x} & 1 \\
		+5_{x} & 1 \\
		+6_{x} & 1 \\
		+7_{x} & 1 
	} \, ,
	\label{eq:cyclesKinesin}
\end{equation}
which entails the intake of two $\ce{ATP}$ molecules and the release of two $\ce{ADP}$ ones
\begin{equation}
	M_{x} := \excMtx_{x} \, C_{x} =
	\kbordermatrix{
		& c_{x} \\
		(E) 			& 0 \\
		(N,\ce{ATP}) 	& 2 \\
		(N,\ce{ADP}) 	& -2 \\
		(X) 			& 0
	} \, ,
	\label{eq:Mkinesin}
\end{equation}
irrespective of the position $x$.
The full $M$-matrix has thus an infinite-number of columns equal to Eq.~\eqref{eq:Mkinesin}, and its three-dimensional cokernel is spanned by
\begin{subequations}
	\begin{align}
		\bm \ell^{\mathrm{E}} & =
		\kbordermatrix{
			& (E) & (N,\ce{ATP}) & (N,\ce{ADP}) & (X) \\
			& 1 & 0 & 0 & 0
		} \label{eq:energyCLkinesin} \\
		\bm \ell^{\mathrm{N}} & =
		\kbordermatrix{
			& (E) & (N,\ce{ATP}) & (N,\ce{ADP}) & (X) \\
			& 0 & 1 & 1 & 0
		} \label{eq:numberCLkinesin} \\
		\bm \ell^{\mathrm{X}} & =
		\kbordermatrix{
			& (E) & (N,\ce{ATP}) & (N,\ce{ADP}) & (X) \\
			& 0 & 0 & 0 & 1
		} \label{eq:positionCLkinesin} \, ,
	\end{align}
	\label{eq:CLkinesin}
\end{subequations}
which clearly corresponds to the three system quantities, $E_{n}$, $N_{n}$, and $X_{n}$, respectively.
As far as the symmetries are concerned, the intersection between its infinite-dimensional column vector space and its (infinite-dimensional) kernel is one-dimensional, in agreement with the observation that all cycles $\st{c_{x}}$ are physically dependent on one.
In other words, there is an infinity of symmetries and all cycles carry the same cycle affinity 
\begin{equation}
	\mathcal{A} = 2 \beta ( \mu_{\ce{ATP}} - \mu_{\ce{ADP}} ) \, ,
	\label{}
\end{equation}
which is thus regarded as the fundamental one.

To illustrate our EP decomposition, we use $(N,\ce{ATP})$ as set of $\push$, while leaving $\st{(E), (N,\ce{ADP}), (X)}$ as $\rf$.
Guided by Eqs.~\eqref{eq:potential} and \eqref{eq:effectiveIntensive}, the potential reads
\begin{equation}
	\phi_{n} = \omega_{n} + \beta k X_{n} \, ,
	\label{eq:potentialKinesin}
\end{equation}
where
\begin{equation}
	\omega_{n} := S_{n} - \beta E_{n} + \beta \mu_{\ce{ADP}} N_{n} \, ,
	\label{eq:grandcanonicalKinesin}
\end{equation}
is the Massieu potential corresponding to the grand potential.
The fundamental forces, Eq.~\eqref{eq:fundForce}, consist solely of
\begin{equation}
	\mathcal{F}_{(N,\ce{ATP})} = \beta (\mu_{\ce{ATP}} - \mu_{\ce{ADP}}) \, .
	\label{eq:fundForceKinesin}
\end{equation}
The EP along a stochastic trajectory with autonomous protocol, Eq.~\eqref{eq:epAwesome}, is
\begin{equation}
	\Sigma[\trj] = \beta \left( \mu_{\ce{ATP}} - \mu_{\ce{ADP}} \right) \mathcal{I}_{\ce{ATP}}[\trj] + \Delta \Phi[\trj] \, ,
	\label{eq:epKinesin}
\end{equation}
where
\begin{equation}
	\hspace{-1.2em}
	\begin{split}
		&\mathcal{I}_{\ce{ATP}}[\trj] := \int_{0}^{t} \de \tau \, \excMtx^{(N,\ce{ATP})}_{e} \, J^{e}(\tau) \\
		& = {\sum_{x=-\infty}^{\infty}} \int_{0}^{t} \de \tau \, \left[ J^{+4_{x}}(\tau) - J^{-4_{x}}(\tau) + J^{+7_{x}}(\tau) - J^{-7_{x}}(\tau) \right]
	\end{split}
	\label{}
\end{equation}
is the total number of $\ce{ATP}$ molecules flowing into the system, while $\Phi$ is the stochastic Massieu potential related to Eq.~\eqref{eq:potentialKinesin}.
Since there is only one fundamental force, the EP in terms of fundamental affinities reads exactly as Eq.~\eqref{eq:epKinesin}.

To illustrate the finite-time detailed FT, let us imagine a system with a finite number of positions $x = 1, \dots, \nof{\mathrm{x}}$.
The potential \eqref{eq:potentialKinesin} thus defines a physical equilibrium state, Eq.~\eqref{eq:pEq}, achieved when the force is turned off:
$\mu_{\ce{ATP}} = \mu_{\ce{ADP}}$.
At time $0$, the autonomous protocol with $\mu_{\ce{ATP}} \neq \mu_{\ce{ADP}}$ (but with the same $\mu_{\ce{ADP}}$ as at equilibrium) is activated and the system moves far from equilibrium.
Notice that any change of $\mu_{\ce{ATP}}$ leaves $\phi_{n}$ unaltered and the process can be stopped at any time.
Hence, the probability of observing the intake of $\mathcal{I}_{\ce{ATP}}$ $\ce{ATP}$ molecules up to time $t$ satisfies
\begin{equation}
	\frac{P_{t}(\mathcal{I}_{\ce{ATP}})}{P_{t}(-\mathcal{I}_{\ce{ATP}})} = \exp\left\{ \beta \, ( \mu_{\ce{ATP}} - \mu_{\ce{ADP}} ) \mathcal{I}_{\ce{ATP}} \right\} \, ,
	\label{}
\end{equation}
see Eq.~\eqref{eq:dftAwesomeAutonomous}.

To formulate a FT which explicitly counts the number of steps, we have to make a step backward and regard the conservative term $\beta k l$ in the local detailed balance, Eq.~\eqref{eq:ldbKinesin}, as an additional force contribution, rather than as part of the potential one.
Under this condition the EP can be recast into
\begin{equation}
	\hspace{-2em}
	\Sigma[\trj] = \beta \left( \mu_{\ce{ATP}} - \mu_{\ce{ADP}} \right) \mathcal{I}_{\ce{ATP}}[\trj]
	+ \beta k \mathcal{X}[\trj] + \Delta \Omega[\trj] \, ,
	\label{eq:epMotorStepping}
\end{equation}
where
\begin{equation}
	\Omega_{n} = \omega_{n} - \ln p_{n}
	\label{eq:granpotential}
\end{equation}
is the stochastic Massieu potential corresponding to Eq.~\eqref{eq:grandcanonicalKinesin}, while
\begin{equation}
	\mathcal{X}[\trj] := X_{n_{t}} - X_{n_{0}}
	\label{}
\end{equation}
the total distance traveled by the motor.
If the system is initially prepared in the grandcanonical equilibrium state---achieved by turning off both the external force $k$ and the fundamental force $\mathcal{F}_{(N,\ce{ATP})}$---the FT reads
\begin{equation}
	\hspace{-1.8em}
	\frac{P_{t}(\mathcal{I}_{\ce{ATP}},\mathcal{X})}{P_{t}(-\mathcal{I}_{\ce{ATP}},-\mathcal{X})} = \exp\left\{ \beta \, ( \mu_{\ce{ATP}} - \mu_{\ce{ADP}} ) \mathcal{I}_{\ce{ATP}} + \beta k \mathcal{X} \right\} \, .
	\label{eq:FTkinesin}
\end{equation}

\subparagraph*{Tightly coupled model}
As an example of change of network topology, we now consider the tightly coupled description in which the transitions $\st{5,6,7}$ are absent, and the network becomes a one-dimensional chain of states.
Since there are no cycles the whole row space of $\excMtx$ spans the conservation laws, which can thus be written as
\begin{subequations}
	\begin{align}
		\bm \ell^{\mathrm{E}} & =
		\kbordermatrix{
			& (E) & (N,\ce{ATP}) & (N,\ce{ADP}) & (X) \\
			& 1 & 0 & 0 & 0
		} \label{eq:energyCLkinesinT} \\
		\bm \ell^{\ce{ATP}} & =
		\kbordermatrix{
			& (E) & (N,\ce{ATP}) & (N,\ce{ADP}) & (X) \\
			& 0 & 1 & 0 & 0
		} \label{eq:atpCLkinesinT} \\
		\bm \ell^{\ce{ADP}} & =
		\kbordermatrix{
			& (E) & (N,\ce{ATP}) & (N,\ce{ADP}) & (X) \\
			& 0 & 0 & 1 & 0
		} \label{eq:adpCLkinesinT} \\
		\bm \ell^{\mathrm{X}} & =
		\kbordermatrix{
			& (E) & (N,\ce{ATP}) & (N,\ce{ADP}) & (X) \\
			& 0 & 0 & 0 & 1
		} \label{eq:positionCLkinesinT} \, .
	\end{align}
	\label{eq:CLkinesinT}
\end{subequations}
With respect to the previous model, the number of $\ce{ATP}$ and $\ce{ADP}$ molecules are separately conserved quantities, Eqs.~\eqref{eq:atpCLkinesinT} and \eqref{eq:adpCLkinesinT}.
The set of fundamental forces is empty while the potential reads
\begin{equation}
	\hspace{-1em}
	\phi_{n} = S_{n} - \beta \left( E_{n} - \mu_{\ce{ATP}} N^{\ce{ATP}}_{n} - \mu_{\ce{ADP}} N^{\ce{ADP}}_{n} - k X_{n} \right) \, ,
	\label{eq:potentialKinesinT}
\end{equation}
thus making the dissipation equal to
\begin{equation}
	\Sigma[\trj] = \Delta \Phi[\trj] \, .
	\label{}
\end{equation}
Therefore, the change of network topology achieved by removing transitions creating cycles, prevents the reservoirs from creating forces.
The potential will be thus described with the maximum amount of conserved quantities, one for each $y$.

\begin{figure}[t]
	\centering
	\includegraphics[width=.30\textwidth]{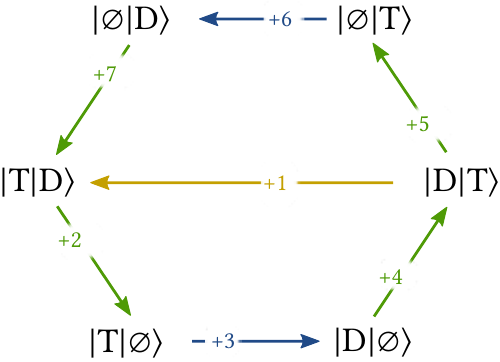}
	\caption{
		Alternative description of the chemomechanical stepping process in Fig.~\ref{fig:networkKinesin}.
		The kinetics and thermodynamics of the internal transitions is unchanged, while the step transition resets the internal motor state.
	}
	\label{fig:periodic}
\end{figure}

\subparagraph*{Alternative Description}
An alternative description of the chemomechanical process is obtained when periodic boundary conditions are imposed, Fig.~\ref{fig:periodic}.
One additional cycle is created,
\begin{equation}
	C =
	\kbordermatrix{
		& c & a \\
		+1 & 0 & 1 \\
		+2 & 1 & 1 \\
		+3 & 1 & 1 \\
		+4 & 1 & 1 \\
		+5 & 1 & 0 \\
		+6 & 1 & 0 \\
		+7 & 1 & 0 
	} \, ,
	\label{eq:cyclesKinesinBC}
\end{equation}
\emph{cf.} Eq.~\eqref{eq:cyclesKinesin}, and the M-matrix now reads
\begin{equation}
	M :=
	\kbordermatrix{
		& \mathrm{c} & \mathrm{a} \\
		(E) 			& 0 & 0 \\
		(N,\ce{ATP}) 	& 2   & 1  \\
		(N,\ce{ADP}) 	& -2  & -1 \\
		(X) 			& 0   & l
	} \, ,
	\label{eq:MkinesinBC}
\end{equation}
As a consequence, the spatial conservation law, \eqref{eq:positionCLkinesin}, is lost and the nonequilibrium Massieu potential becomes $\omega_{n}$, Eqs.~\eqref{eq:grandcanonicalKinesin} and \eqref{eq:granpotential}.
However, the set of fundamental forces gains one element,
\begin{equation}
	\mathcal{F}_{(X)} = \beta k \, ,
	\label{}
\end{equation}
which is conjugated to the traveled distance:
\begin{equation}
	\begin{split}
		\mathcal{X}[\trj] & := \int_{0}^{t} \de \tau \, \excMtx^{(X)}_{e} \, J^{e}(\tau) \\
		& = l \int_{0}^{t} \de \tau \, \left[ J^{+1_{x}}(\tau) - J^{-1_{x}}(\tau) \right] \, .
	\end{split}
	\label{}
\end{equation}
Hence, the expression of the EP and the formulation of the finite-time detailed FT read as in Eqs.~\eqref{eq:epMotorStepping} and \eqref{eq:FTkinesin}, respectively.

In conclusion, the periodic boundary condition can be viewed as a change of network topology in which one conservation law is lost and a fundamental force emerges.

\subsection{Randomized Grid}
\label{sec:REG}

As a final illustration, we consider a particle hopping between states positioned at the nodes of a two-dimensional grid, $n=(x,z)$ for $x,z=1,\dots,\nof{}$.
The transitions along the edges are triggered by randomly distributed work reservoirs.
This model provides an example of systems which could not be analyzed thermodynamically without resorting to our systematic procedure.
It also shows how physical topological alterations may give rise to symmetry changes which in turn affects the thermodynamics.

The states are characterized by a spatial coordinate $X_{n} = a_{\mathrm{x}} x + a_{\mathrm{z}} z$, and jumps are only allowed between nearest neighbors:
$x \rightarrow x \pm 1$ or $z \rightarrow z \pm 1$.
The system is isothermal and each transition is ruled by a force $f_{(X,r)} = - \beta k_{r}$, which is initially drawn randomly from a set of $\nof{\mathrm{r}}$ reservoirs.
The $\excMtx$-matrix relating transitions to reservoirs is given by
\begin{equation}
	\excMtx^{r}_{e} =
	\begin{cases}
		\pm a_{\mathrm{x}} & \text{if } e = x \overset{r}{\longrightarrow} x \pm 1 \\
		\pm a_{\mathrm{z}} & \text{if } e = z \overset{r}{\longrightarrow} z \pm 1 \\
		  0 & \text{otherwise}
	\end{cases} \, .
	\label{eq:Ygrid}
\end{equation}
\emph{i.e.} if $e$ is triggered by the work reservoir $r$, then $\excMtx^{r}_{e}$ is equal to $\pm a_{\mathrm{x}}$ or $\pm a_{\mathrm{z}}$ depending on the direction of the transition.

\begin{figure}[t]
	\centering
	\includegraphics[width=.30\textwidth]{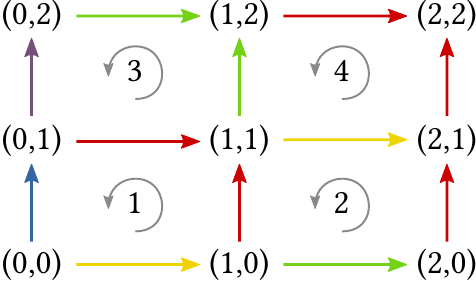}
	\caption{
		Illustration of a 3$\times$3 grid with nearest-neighbor transitions triggered by a reservoir chosen at random among five.
		The color of each transition corresponds to a different reservoir:
		1, yellow; 2, green; 3, purple; 4, blue; and 5, red.
	}
	\label{fig:randomGrid}
\end{figure}

As an example, we consider the 3$\times$3 grid coupled to 5 reservoirs depicted in Fig.~\ref{fig:randomGrid}.
We omit to report the matrices $\excMtx$ and $C$ as they can be easily inferred form Eq.~\eqref{eq:Ygrid} and the picture, and move on to the $M$-matrix, which reads
\begin{equation}
	M =
	\kbordermatrix{
		& 1 & 2 & 3 & 4 \\
		(X,1) & a_{\mathrm{x}} & -a_{\mathrm{x}} & 0 & a_{\mathrm{x}} \\
		(X,2) & 0 & a_{\mathrm{x}} & a_{\mathrm{z}}-a_{\mathrm{x}} & -a_{\mathrm{z}} \\
		(X,3) & 0 & 0 & -a_{\mathrm{z}} & 0 \\
		(X,4) & -a_{\mathrm{z}} & 0 & 0 & 0 \\
		(X,5) & a_{\mathrm{z}}-a_{\mathrm{x}} & 0 & a_{\mathrm{x}} & a_{\mathrm{z}}-a_{\mathrm{x}}
	} \, .
	\label{}
\end{equation}
Its one-dimensional cokernel is spanned by the vector
\begin{equation}
	\bm \ell^{\mathrm{X}} =
	\kbordermatrix{
		& (X,1) & (X,2) & (X,3) & (X,4) & (X,5) \\
		& 1 & 1 & 1 & 1 & 1
	} \\
	\label{}
\end{equation}
which corresponds to the global conserved quantity $X_{n}$.
In contrast, its kernel is empty denoting the absence of symmetries.
Setting $-\beta k_{1}$ as ``potential'' field, $\rf$, the nonequilibrium potential reads
\begin{equation}
	\phi_{n} = \beta k_{1} X_{n} \, ,
	\label{}
\end{equation}
while the fundamental forces are equal to
\begin{equation}
	\mathcal{F}_{(X,r)} = \beta \left( k_{r} - k_{1} \right) \, , \quad \text{for } r = 2, \dots, 5 \, .
	\label{}
\end{equation}
The trajectory EP can be thus expressed as
\begin{equation}
	\Sigma[\trj] = v[\trj] + \sum_{r=2}^{5} \sigma_{r}[\trj] + \Delta \Phi[\trj] \, ,
	\label{}
\end{equation}
where
\begin{subequations}
	\begin{align}
		v[\trj] &:= - \beta \int_{0}^{t} \de \tau \, \partial_{\tau} \at{\left[ k_{1}(\tau) X_{n}(\tau) \right]}{n=n_{\tau}} \\
		\sigma_{r}[\trj] &:= \beta \int_{0}^{t} \de \tau \, \left[ k_{r}(\tau) - k_{1}(\tau) \right] I_{r}(\tau) \, .
	\end{align}
\end{subequations}

\begin{figure}[t]
	\centering
	\includegraphics[width=.26\textwidth]{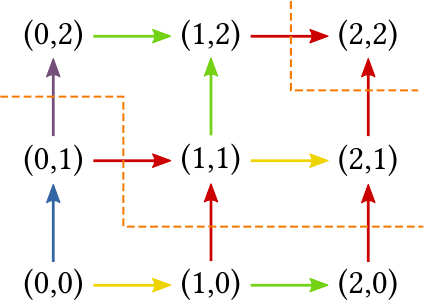}
	\caption{
		Illustration of the randomized grid in Fig.~\ref{fig:randomGrid} for $a_{\mathrm{x}} = a_{\mathrm{z}} = a$.
		The grid is split into three groups of states by the transitions corresponding to the third (purple) and fifth (red) reservoir:
		$\st{(0,0), (1,0), (0,1), (2,0)}$, $\st{(1,1), (0,2), (2,1), (1,2)}$, and $\st{(2,2)}$.
	}
	\label{fig:RGDB}
\end{figure}

In order to show the emergence of a symmetry following a change of physical topology, let us now assume that $a_{\mathrm{x}} = a_{\mathrm{z}} = a$ and carry on the same analysis as before.
The $M$-matrix now becomes,
\begin{equation}
	M =
	\kbordermatrix{
		& 1 & 2 & 3 & 4 \\
		(X,1) & a & -a & 0 & a \\
		(X,2) & 0 & a & 0 & -a \\
		(X,3) & 0 & 0 & -a & 0 \\
		(X,4) & -a & 0 & 0 & 0 \\
		(X,5) & 0 & 0 & a & 0
	} \, .
	\label{}
\end{equation}
whose kernel and cokernel are one and two-dimensional, respectively.
The symmetries are given by
\begin{equation}
	\sym = 
		\kbordermatrix{
			& 1 & 2 & 3 & 4 \\
			& 0 & 1 & 0 & 1
		} \, ,
	\label{}
\end{equation}
and tell us that the second and fourth cycles are not physically independent, as they are coupled to the same reservoirs and all displacements are the same.
The basis of $\coker M$,
\begin{subequations}
	\begin{align}
		\bm \ell^{\mathrm{X}} & =
		\kbordermatrix{
			& (X,1) & (X,2) & (X,3) & (X,4) & (X,5) \\
			& 1 & 1 & 1 & 1 & 1
		} \\
		\bm \ell^{V} & =
		\kbordermatrix{
			& (X,1) & (X,2) & (X,3) & (X,4) & (X,5) \\
			& 0 & 0 & 1 & 0 & 1
		}
	\end{align}
	\label{eq:}
\end{subequations}
identifies two state variables, the first of which is the global conserved quantity, $X_{n}$, whereas the second is
\begin{equation}
	\hspace{-1.2em}
	V_{n} = 
	\kbordermatrix{
		& (0,0) & (1,0) & (0,1) & (2,0) & (1,1) & (0,2) & (2,1) & (1,2) & (2,2) \\
		& 0 & 0 & 0 & 0 & a & a & a & a & 2 a
	}
	\label{eq:XRG}
\end{equation}
whose interpretation is not obvious.
It arises from the fact that $x$- and $z$-transitions are indistinguishable and the reservoirs $3$ and $5$ split the states into three groups, see Fig.~\ref{fig:RGDB}, which are identified by different values of $V_{n}$, Eq.~\eqref{eq:XRG}.
We can set $(X,1)$ and $(X,3)$ as the reservoirs of the set $\st{\rf}$, according to which the Massieu potential of the state $n$ reads
\begin{equation}
	\phi_{n} = \beta \left[ k_{1} X_{n} + ( k_{3} - k_{1} ) V_{n} \right] \, .
	\label{eq:potentialRG}
\end{equation}
The number of fundamental forces is thus reduced,
\begin{subequations}
	\begin{align}
		\mathcal{F}_{(X,2)} & = \beta k_{2} - \beta k_{1} \, , \\
		\mathcal{F}_{(X,4)} & = \beta k_{4} - \beta k_{1} \, , \\
		\mathcal{F}_{(X,5)} & = \beta k_{5} - \beta k_{3} \, .
	\end{align}
	\label{eq:fundamentalRG}
\end{subequations}
The EP can be easily written.

This model exemplifies the emergence of nontrivial conservation laws whose identification is not straightforward, and motivates the need for a systematic procedure capable of separating the conservative contributions to the EP from the nonconservative ones.

\section{Conclusions and Perspectives}

The central achievement of this paper is to show that the EP of an open system described by stochastic thermodynamics is shaped by the way conserved quantities constrain the exchanges between the system and the reservoirs.
Some of these conserved quantities are the obvious ones which do not depend on the system details (\emph{e.g.} energy, particle number).
But we provide a systematic procedure to identify the nontrivial ones which depend on the system topology.
As a result, we can split the EP into three fundamental contributions, one solely caused by the time-dependent drivings, another expressed as the change of a nonequilibrium Massieu potential, and a third one which contains the fundamental set of flux and forces.
Table \ref{tab:EPprocesses} indicates which of these contributions play a role in different known processes.
We also showed how to make use of this decomposition to derive a finite-time detailed FT solely expressed in terms of physical quantities, as well as to asses the cost of manipulating nonequilibrium states via time-dependent driving and nonconservative forces.

We believe that this work provides a comprehensive formulation of stochastic thermodynamics.
Our framework can be systematically used to study any specific model (as we illustrated on several examples) and demonstrates the crucial importance of conservation laws in thermodynamics, at, as well as out of, equilibrium.

\begin{acknowledgments}
	We thank G.~Bulnes Cuetara for advises on the FT proof.
	This work was funded by the National Research Fund of Luxembourg (AFR PhD Grant 2014-2, No.~9114110) and the European Research Council project NanoThermo (ERC-2015-CoG Agreement No. 681456). 
\end{acknowledgments}

\appendix

\section{Proof of the one-to-one correspondence between fundamental forces and fundamental affinities}
\label{sec:proofNonsingularity}

We need to prove that that the matrix whose entries are $\st{M^{\push}_{\eta}}$ is nonsingular given the following hypotheses:
\emph{(i)} the vectors labeled by $\eta$ whose entries are $\st{M^{y}_{\eta}}$, for $y=1,\dots,\nof{\mathrm{y}}$, are linearly independent;
\emph{(ii)} $\ell^{\lambda}_{\push} M^{\push}_{\alpha} + \ell^{\lambda}_{\rf} M^{\rf}_{\alpha} = 0$ for all $\lambda$ and $\alpha$, where the matrix whose entries are $\st{\ell^{\lambda}_{\push}}$ is nonsingular.
Let us now assume by contradiction that $\st{M^{\push}_{\eta}}$ is singular, and let us denote by $\st{x^{\eta}}$ the entries of a non-null vector such that $M^{\push}_{\eta} x^{\eta} = 0$ for all $\push$.
We can thus construct a vector $\st{x^{\alpha}}$ having as entries corresponding to $\eta$, $\st{x^{\eta}}$, and zero for the others.
Hence, $M^{\push}_{\alpha} x^{\alpha} = 0$ for all $\push$.
From the equation in the second hypothesis, we get
\begin{equation*}
	\ell^{\lambda}_{\rf} M^{\rf}_{\alpha} x^{\alpha} + \ell^{\lambda}_{\push} M^{\push}_{\alpha} x^{\alpha} = \ell^{\lambda}_{\rf} M^{\rf}_{\eta} x^{\eta} = 0 \, .
\end{equation*}
Since the matrix whose entries are $\st{\ell^{\lambda}_{\rf}}$ is nonsingular, we must conclude that $M^{\rf}_{\eta} x^{\eta} = 0$ for all $\rf$, and thus $M^{y}_{\eta} x^{\eta} = 0$ for all $y$, in contradiction with the hypothesis \emph{(i)}.

\section{Proof of the finite-time detailed FTs}
\label{sec:proof}

We now give the proof of the finite time detailed FTs \eqref{eq:dftAwesome} using moment generating functions.
Alternatively, it can be proved using the approach developed in Ref.~\cite{garcia-garcia10}.
For our purposes, we change our notation for a bracket operatorial one.

Let ${P_{t}(n,v,\st{\sigma_{\push}})}$ be the joint probability of observing a trajectory ending in the state $n$ along which the driving contribution is $v$ while the flow ones are $\st{\sigma_{\push}}$.
The above probabilities, one for each $n$, are stacked in the ket $\ket{P_{t}(v, \st{\sigma_{\push}})}$.
The time evolution of the moment generating function of the above probabilities,
\begin{multline}
	\ket{\Lambda_{t} (\xi_{\mathrm{d}},\st{\xi_{\push}})} := \int \de v {\textstyle \prod_{\push}} \de \sigma_{\push} \\
	\exp\left\{ -\xi_{\mathrm{d}} v - \xi^{\push} \sigma_{\push} \right\} \ket{P_{t}(v, \st{\sigma_{\push}})} \, ,
	\label{eq:mgfnApp}
\end{multline}
is ruled by the biased stochastic dynamics
\begin{equation}
	\dt \ket{\Lambda_{t}(\xi_{\mathrm{d}},\st{\xi_{\push}})} = {\mathcal{W}}_{t}(\xi_{\mathrm{d}},\st{\xi_{\push}}) \ket{\Lambda_{t}(\xi_{\mathrm{d}},\st{\xi_{\push}})} \, ,
	\label{eq:biasedME}
\end{equation}
where the entries of the biased generator are given by
\begin{multline}
	\mathcal{W}_{nm,t}(\xi_{\mathrm{d}},\st{\xi_{\push}}) \\
	= {\textstyle\sum_{e}} w_{e} \big\{ \exp\left\{ - \xi^{\push} \mathcal{F}_{\push} \excMtx_{\push,e} \right\} \delta_{n,o(-e)} \delta_{m,o(e)} \\
	+ \delta_{n,m} \delta_{m,o(e)} \big\} + \xi_{\mathrm{d}} \partial_{t} \phi_{m} \delta_{n,m} \, .
	\label{eq:biasedGeneratorBis}
\end{multline}
Because of the local detailed balance \eqref{eq:ldbAwesome}, the stochastic generator satisfies the following symmetry
\begin{equation}
	{\mathcal{W}}_{t}\transpose(\xi_{\mathrm{d}},\st{\xi_{\push}}) = \mathcal{B}_{t}^{-1} \, \mathcal{W}_{t}(\xi_{\mathrm{d}},\st{ 1 - \xi_{\push} }) \, \mathcal{B}_{t} \, ,
	\label{eq:symmetryOper}
\end{equation}
where the entries of $\mathcal{B}_{t}$ are given by
\begin{equation}
	\mathcal{B}_{nm,t} := \exp\left\{\phi_{m} \right\} \delta_{n,m} \, .
	\label{eq:B}
\end{equation}
Also, the initial condition is given by the equilibrium distribution \eqref{eq:pEq}, which reads
\begin{equation}
	\ket{\Lambda_{0}(\xi_{\mathrm{d}},\st{\xi_{\push}})} = \ket{p_{\mathrm{eq}_{\mathrm{i}}}} = \mathcal{B}_{0}/Z_{0} \ket{1} \, ,
	\label{}
\end{equation}
where $Z_{0} := \exp\{ \Phi_{\mathrm{eq}_{\mathrm{i}}} \}$ is the partition function.
The ket $\ket{1}$ refers to the vector in the state space whose entries are all equal to one.

In order to proceed further, it is convenient to first prove a preliminary result.
Let us consider the generic biased dynamics, \emph{e.g.} Eq.~\eqref{eq:biasedME},
\begin{equation}
	\dt \ket{\Lambda_{t}(\xi)} = \mathcal{W}_{t}(\xi) \ket{\Lambda_{t}(\xi)} \, ,
	\label{eq:generation}
\end{equation}
whose initial condition is $\ket{\Lambda_{0}(\xi)} = \ket{p(0)}$.
A formal solution of Eq.~\eqref{eq:generation} is $\ket{\Lambda_{t}(\xi)} = {\mathcal{U}}_{t}(\xi) \, \ket{p(0)}$, where the time-evolution operator reads $\mathcal{U}_{t}(\xi) = \mathcal{T}_{+} \exp\left\{ \int_{0}^{t} \de \tau \, {\mathcal{W}}_{\tau}(\xi) \right\}$, $\mathcal{T}_{+}$ being the time-ordering operator.
We clearly have $\dt \mathcal{U}_{t}(\xi) = {\mathcal{W}}_{t}(\xi) \mathcal{U}_{t}(\xi)$.
Let us now consider the following transformed evolution operator
\begin{equation}
	\tilde{\mathcal{U}}_{t}(\xi) := \mathcal{X}^{-1}_{t} \mathcal{U}_{t}(\xi) {\mathcal{X}}_{0} \, ,
	\label{eq:transformedUdef}
\end{equation}
${\mathcal{X}}_{t}$ being a generic invertible operator.
Its dynamics is ruled by the following biased stochastic dynamics
\begin{equation}
	\begin{split}
		\dt \tilde{\mathcal{U}}_{t}(\xi) & = \dt \mathcal{X}^{-1}_{t} \mathcal{U}_{t}(\xi) {\mathcal{X}}_{0} + \mathcal{X}^{-1}_{t} \dt \mathcal{U}_{t}(\xi) {\mathcal{X}}_{0} \\
		& = \left\{ \dt \mathcal{X}^{-1}_{t} {\mathcal{X}}_{t} + \mathcal{X}^{-1}_{t} {\mathcal{W}}_{t}(\xi) {\mathcal{X}}_{t} \right\} \tilde{\mathcal{U}}_{t}(\xi) \\
		& \equiv \tilde{\mathcal{W}}_{t}(\xi) \, \tilde{\mathcal{U}}_{t}(\xi) \, ,
	\end{split}
	\label{eq:transformedUdynamics}
\end{equation}
which allows us to conclude that the transformed time-evolution operator is given by
\begin{equation}
	\tilde{\mathcal{U}}(\xi) = \mathcal{T}_{+} \exp\left\{ \int_{0}^{t} \de \tau \, \tilde{\mathcal{W}}_{\tau}(\xi) \right\} \, .
	\label{eq:transformedU}
\end{equation}

\begin{widetext}
From Eqs.~\eqref{eq:transformedUdef}, \eqref{eq:transformedUdynamics} and \eqref{eq:transformedU} we deduce that
\begin{equation}
	\mathcal{X}^{-1}_{t} \mathcal{U}_{t}(\xi) {\mathcal{X}}_{0} = \mathcal{T}_{+} \exp\left\{ \int_{0}^{t} \de \tau \, \left[ \de_{\tau} \mathcal{X}^{-1}_{\tau} {\mathcal{X}}_{\tau} + \mathcal{X}^{-1}_{\tau} {\mathcal{W}}_{\tau}(\xi) {\mathcal{X}}_{\tau} \right] \right\} \, .
	\label{eq:magic}
\end{equation}

We can now come back to our specific biased stochastic dynamics \eqref{eq:biasedME}.
The moment generating function of ${P_{t}(v, \st{\sigma_{\push}})}$ is thus given by
\begin{equation}
	\Lambda_{t}(\xi_{\mathrm{d}},\st{\xi_{\push}}) 
	=  \braket{1|\Lambda_{t}(\xi_{\mathrm{d}},\st{\xi_{\push}})} 
	= \braket{1| \mathcal{U}_{t}(\xi_{\mathrm{d}},\st{\xi_{\push}}) \mathcal{B}_{0} / Z_{0} | 1}
	= \braket{1|\frac{\mathcal{B}_{t}}{Z_{t}} \mathcal{B}_{t}^{-1} \, \mathcal{U}_{t}(\xi_{\mathrm{d}},\st{\xi_{\push}}) \, \mathcal{B}_{0} | 1} \frac{Z_{t}}{Z_{0}} \, ,
	\label{eq:proofFirst}
\end{equation}
where $\mathcal{U}_{t}(\xi_{\mathrm{d}},\st{\xi_{\push}})$ is the time-evolution operator of the biased stochastic dynamics \eqref{eq:biasedME}.
The requirement imposed on $\pi_{t}$---discussed in the main text---ensures that $\bra{1}{\mathcal{B}_{t}}/{Z_{t}}$ with $Z_{t}:=\exp\{ \Phi_{\mathrm{eq}_{\mathrm{f}}} \}$ is the equilibrium initial distribution of the backward process $\bra{p_{\mathrm{eq}_{\mathrm{f}}}}$.
Using the relation in Eq.~\eqref{eq:magic}, the above term can be rewritten as
\begin{equation}
	= \braket{p_{\mathrm{eq}_{\mathrm{f}}} | \mathcal{T}_{+} \exp\left\{ \int_{0}^{t} \de \tau \, \left[ \partial_{\tau} \mathcal{B}_{\tau}^{-1} \mathcal{B}_{\tau} + \mathcal{B}_{\tau}^{-1} \, {\mathcal{W}}_{\tau}(\xi_{\mathrm{d}},\st{\xi_{\push}}) \, \mathcal{B}_{\tau} \right] \right\} |1} \exp\left\{ \Delta \Phi_{\mathrm{eq}} \right\} \, ,
	\label{}
\end{equation}
where $\Delta \Phi_{\mathrm{eq}} \equiv \ln Z_{\mathrm{t}}/Z_{0}$.
Since $\partial_{\tau} \mathcal{B}_{\tau}^{-1} \mathcal{B}_{\tau} = \diag\left\{ - \partial_{\tau} \phi_{n} \right\}$ the first term in square bracket can be added to the diagonal entries of the second term, thus giving
\begin{equation}
	= \braket{p_{\mathrm{eq}_{\mathrm{f}}} | \mathcal{T}_{+} \exp\left\{ \int_{0}^{t} \de \tau \, \left[ \mathcal{B}_{\tau}^{-1} \, {\mathcal{W}}_{\tau}(\xi_{\mathrm{d}} - 1,\st{\xi_{\push}}) \, \mathcal{B}_{\tau} \right] \right\} |1} \exp\left\{ \Delta \Phi_{\mathrm{eq}} \right\} \, .
	\label{}
\end{equation}
The symmetry \eqref{eq:symmetryOper} allow us to recast the latter into
\begin{equation}
	= \braket{p_{\mathrm{eq}_{\mathrm{f}}} | \mathcal{T}_{+} \exp\left\{ \int_{0}^{t} \de \tau \, \mathcal{W}\transpose_{\tau}\left( \xi_{\mathrm{d}} - 1, \st{ 1 - \xi_{\push} } \right) \right\} |1} \exp\left\{ \Delta \Phi_{\mathrm{eq}} \right\} \, .
	\label{eq:intermediario}
\end{equation}
The crucial step comes as we transform the integration variable from $\tau$ to $\tau^{\dagger} = t - \tau$.
Accordingly, the time-ordering operator, $\mathcal{T}_{+} $, becomes an anti-time-ordering one $\mathcal{T}_{-}$, while the diagonal entries of the biased generator become
\begin{equation}
	\begin{split}
		{\mathcal{W}}_{mm,t-\tau^{\dagger}}(\xi_{\mathrm{d}},\st{\xi_{\push}})
		& = {\textstyle\sum_{e}} w_{e}(t-\tau^{\dagger}) \, \delta_{m,o(e)} + \xi_{\mathrm{d}} \, \partial_{(t-\tau^{\dagger})} \phi_{m}(t-\tau^{\dagger}) \\
		& = {\textstyle\sum_{e}} w_{e}(t-\tau^{\dagger}) \, \delta_{m,o(e)} - \xi_{\mathrm{d}} \, \partial_{\tau^{\dagger}} \phi_{m}(t-\tau^{\dagger}) \, ,
	\end{split}
	\label{}
\end{equation}
from which we conclude that
\begin{equation}
		{\mathcal{W}}_{nm,t-\tau^{\dagger}}(\xi_{\mathrm{d}},\st{\xi_{\push}})
		= {\mathcal{W}}_{nm,t-\tau^{\dagger}}(- \xi_{\mathrm{d}},\st{\xi_{\push}})
		=: {\mathcal{W}}^{\dagger}_{nm,\tau^{\dagger}}(-\xi_{\mathrm{d}},\st{\xi_{\push}}) \, .
	\label{}
\end{equation}
Above, ${\mathcal{W}}^{\dagger}_{\tau^{\dagger}}(\xi_{\mathrm{d}},\st{\xi_{\push}})$ is the biased generator of the dynamics subject to the time-reversed protocol, $\pi^{\dagger}$, \emph{i.e.} the dynamics of the backward process.
Equation~\eqref{eq:intermediario} thus becomes
\begin{equation}
	= \braket{p_{\mathrm{eq}_{\mathrm{f}}} | \mathcal{T}_{-} \exp\left\{ \int_{0}^{t} \de \tau^{\dagger} \, {\mathcal{W}_{\tau^{\dagger}}^{\dagger}}\transpose \left( 1 - \xi_{\mathrm{d}},\st{ 1 - \xi_{\push} } \right) \right\} |1} \exp\left\{ \Delta \Phi_{\mathrm{eq}} \right\} \, .
	\label{}
\end{equation}
Upon a global transposition, we can write
\begin{equation}
	= \braket{1 | \mathcal{T}_{+} \exp\left\{ \int_{0}^{t} \de \tau^{\dagger} \, {\mathcal{W}_{\tau^{\dagger}}^{\dagger}} \left( 1 - \xi_{\mathrm{d}},\st{ 1 - \xi_{\push} } \right) \right\} |p_{\mathrm{eq}_{\mathrm{f}}}} \exp\left\{ \Delta \Phi_{\mathrm{eq}} \right\} \, ,
\end{equation}
\end{widetext}
where we also used the relationship between transposition and time-ordering
\begin{equation}
	\mathcal{T}_{+} \left( {\textstyle\prod_{i}} A\transpose_{t_{i}} \right) = \left( \mathcal{T}_{-} {\textstyle\prod_{i}} A_{t_{i}} \right)\transpose \, ,
	\label{}
\end{equation}
in which $A_{t}$ is a generic operator.
From the last expression, we readily obtain
\begin{equation}
	\begin{split}
		& = \braket{1 | {\mathcal{U}^{\dagger}_{t}} \left( 1 - \xi_{\mathrm{d}},\st{ 1 - \xi_{\push} } \right) |p_{\mathrm{eq}_{\mathrm{f}}}} \exp\left\{ \Delta \Phi_{\mathrm{eq}} \right\} \\
		& = \Lambda^{\dagger}_{t}\left( 1 - \xi_{\mathrm{d}},\st{ 1 - \xi_{\push} } \right)  \, \exp\left\{ \Delta \Phi_{\mathrm{eq}} \right\} \, ,
	\end{split}
	\label{}
\end{equation}
where $\Lambda^{\dagger}_{t}\left( \xi_{\mathrm{d}},\st{ \xi_{\push} } \right)$ is the moment generating function of $P^{\dagger}_{t}(v, \st{\sigma_{\push}})$.
Summarizing, we have the following symmetry
\begin{equation}
	\hspace{-1em}
	\Lambda_{t} (\xi_{\mathrm{d}},\st{\xi_{\push}}) = \Lambda^{\dagger}_{t}\left( 1 - \xi_{\mathrm{d}},\st{ 1 - \xi_{\push} } \right)  \, \exp\left\{ \Delta \Phi_{\mathrm{eq}} \right\} \, ,
	\label{eq:proofLast}
\end{equation}
whose inverse Laplace transform gives the FT
\begin{equation}
	\frac{P_{t}(v, \st{\sigma_{\push}})}{P^{\dagger}_{t}(-v, \st{-\sigma_{\push}})} = 
	\exp \left\{ v + {\textstyle\sum_{\push}} \sigma_{\push} + \Delta \Phi_{\mathrm{eq}} \right\} \, .
\end{equation}

\subsection*{Fundamental Cycles}
\label{sec:proofCycles}

The finite-time detailed FT for flow contributions along fundamental cycles, Eq.~\eqref{eq:dftSuperAwesome}, follows the same logic and mathematical steps described above.
The moment generating function which now must be taken into account is
\begin{multline}
	\ket{\Lambda_{t} (\xi_{\mathrm{d}},\st{\xi_{\eta}})} := \int \de v {\textstyle \prod_{\eta}} \de \gamma_{\eta} \\
	\exp\left\{ -\xi_{\mathrm{d}} v - \xi^{\eta} \gamma_{\eta} \right\} \ket{P_{t}(v, \st{\gamma_{\eta}})} \, ,
	\label{eq:mgfnCycles}
\end{multline}
which is ruled by the biased generator whose entries are
\begin{multline}
	\mathcal{W}_{nm,t}(\xi_{\mathrm{d}},\st{\xi_{\eta}}) \\
	= {\textstyle\sum_{e}} w_{e} \big\{ \exp\left\{ - \xi^{\eta} \mathcal{A}_{\eta} \zeta_{\eta,e} \right\} \delta_{n,o(-e)} \delta_{m,o(e)} \\
	+ \delta_{n,m} \delta_{m,o(e)} \big\} + \xi_{\mathrm{d}} \partial_{t} \phi_{m} \delta_{n,m} \, .
	\label{eq:biasedGeneratorAff}
\end{multline}
The symmetry of the latter generator---on top of which the proof is constructed---is based on the expression of the local detailed balance given in Eq.~\eqref{eq:ldbAwesome},
\begin{equation}
	{\mathcal{W}}_{t}\transpose(\xi_{\mathrm{d}},\st{\xi_{\eta}}) = \mathcal{B}_{t}^{-1} \, \mathcal{W}_{t}(\xi_{\mathrm{d}},\st{ 1 - \xi_{\eta} }) \, \mathcal{B}_{t} \, ,
	\label{eq:symmetryOperAff}
\end{equation}
where the entries of $\mathcal{B}_{t}$ are given in Eq.~\eqref{eq:B}.
Following the steps from Eq.~\eqref{eq:proofFirst} to Eq.~\eqref{eq:proofLast}, with the above definitions and equations, Eqs.~\eqref{eq:mgfnCycles}--\eqref{eq:symmetryOperAff}, proves the FT in Eq.~\eqref{eq:dftSuperAwesome}.

\bibliography{indiceLocale}

\end{document}